\documentclass[12pt,preprint]{aastex}
\usepackage{natbib}
\usepackage{graphicx}
\begin{document}
\title{Near-infrared spectroscopy of five Blue Compact Dwarf galaxies:
II~Zw~40, Mrk~71, Mrk~930, Mrk~996 and SBS~0335$-$052E\footnote{Tables 
\ref{tab3} -- \ref{tab7} are presented only in the electronic edition of the 
{\sl Journal}.}}
\author{Yuri I. Izotov}
\affil{Main Astronomical Observatory, Ukrainian National Academy of Sciences,
27 Zabolotnoho str., Kyiv 03680, Ukraine}
\email{izotov@mao.kiev.ua}
\and
\author{Trinh X. Thuan}
\affil{Astronomy Department, University of Virginia, P.O. Box 400325, 
Charlottesville, VA 22904-4325}
\email{txt@virginia.edu}

\begin{abstract}
We present near-infrared (NIR) spectroscopic observations 
of five blue compact dwarf 
(BCD) galaxies, II Zw 40, Mrk 71, Mrk 930, Mrk 996 and SBS 0335--052E.   
The NIR spectra which cover the 0.90 $\mu$m -- 2.40 $\mu$m
wavelength range, show hydrogen, 
molecular hydrogen, helium, sulfur and iron emission
lines. The NIR data for all BCDs have been supplemented by 
optical spectra. We found the extinction coefficient
in all BCDs to be very similar in both the optical and NIR ranges.   
The NIR hydrogen emission lines do not reveal more   
star formation than seen in the optical. 
The same conclusion is reached from Spitzer data 
concerning the MIR emission lines.   
This implies that emission-line spectra
of low-metallicity BCDs in the $\sim$ 0.36 -- 25 $\mu$m wavelength range 
are emitted by relatively transparent
ionized gas. The large extinction derived from the MIR continuum emission 
in some BCDs implies that the latter
arises not from the visible H {\sc ii} regions themselves, but from 
locations outside these H {\sc ii} regions.
The  H$_2$ emission line fluxes can be accounted for by fluorescence. 
CLOUDY stellar photoinization models of all BCDs 
reproduce well the 
fluxes of most of the observed optical and NIR emission lines, 
except in Mrk 930 where shock ionization is needed to account for 
the [Fe {\sc ii}] emission lines. However, some contribution
of shock ionization at the level of $\la$ 10\% that of stellar
ionization is required to reproduce the observed fluxes of high ionization
species, such as He {\sc ii} $\lambda$0.469 $\mu$m in the 
optical range and [O {\sc iv}] $\lambda$25.89 $\mu$m in the MIR range.
\end{abstract}


\section{INTRODUCTION\label{sec:INT}}

Blue compact dwarf (BCD) galaxies are the least chemically 
evolved star-forming galaxies known in the universe 
\citep[see ][for a review]{thuan08}. They have an oxygen abundance 
in the range 12 + log O/H = 7.1 -- 8.3, i.e. 1/40 
--1/3 that of the Sun if the solar abundance of 
\citet{asplund09}, 12 + log O/H = 8.7, is adopted. 
They thus constitute excellent
nearby laboratories for studying star formation and 
its interaction with the interstellar medium (ISM) in a very metal-deficient
environment, intermediate between the solar-metallicity 
environment of galaxies such as the
Milky Way, and the pristine environment of primordial galaxies.  
While BCDs have been extensively studied spectroscopically   
in the optical range, not 
much work has been done in the near-infrared (NIR) 0.9 -- 2.4 $\mu$m range. 
We have thus started a long-term observational program to study 
the NIR spectroscopic properties of a sample of 
BCDs chosen to cover a large range of metallicities,
with 12 + log O/H ranging from 7.1 to 8.3. The 
first results of the program, concerning the 
BCD Mrk 59 (12 + log O/H =  8.0), have been discussed by \citet{I09}.   
Here, we present new  $J$, $H$ and $K$ spectroscopic observations 
of five more BCDs, with 12 + log O/H ranging from 7.29 to 8.10. 

In this new list, only two BCDs have had previous NIR observations.
II Zw 40 has been studied by \citet{JL88}, \citet{MO88}, \citet{M90},
\citet{V96}, \citet{C01} and \citet{V08}, while SBS 0335--052E
has been observed by \citet{V00}. However, all these observations
covered a more restricted wavelength range, including 
only either the $H$ band and/or the $K$ band.
The other three BCDs, Mrk 71, Mrk 930 and 
Mrk 996, have not been observed spectroscopically in the NIR before. 
The general characteristics of the five objects 
are given in Table \ref{tab1}. 
With advances in NIR detector sensitivities, 
and with the commisionning in 2008 of the TripleSpec spectrograph at the 
Apache Point Observatory (APO) 3.5-meter telescope\footnote{The Apache Point 
Observatory 
3.5-meter telescope is owned and operated by the Astrophysical Research 
Consortium.}, we can obtain NIR spectra 
with a simultaneous coverage of the $JHK$ bands
 and considerably more 
signal-to-noise. These improved observations will  
allow us to study in more detail the physical conditions in BCDs, 
and check for their dependence on metallicity.
We will be able to compare the   
extinction in the optical and NIR and search for hidden star formation. 
We will also be able to study the excitation mechanisms
of line emission of ionized species in the H {\sc ii} region, and 
search for molecular hydrogen emission.

In addition to the NIR observations, we have also obtained optical spectra 
of the same objects, except for Mrk 930, in
the wavelength range 0.36 -- 0.97 $\mu$m. These were taken so to 
have an overlap with 
the NIR spectra in the 0.90 -- 0.97 $\mu$m range, 
which allows us to match the two types of spectra and  reduce the  
uncertainties due to 
different apertures and/or varying weather conditions. 

We describe the 3.5m APO observations for the 
five new BCDs in Section \ref{sec:OBS}. In Section \ref{sec:RES}, 
we discuss their extinction, both in the optical and NIR ranges, and 
in the MIR range using Spitzer data, 
the excitation mechanisms for 
molecular hydrogen and iron emission in the NIR range, and CLOUDY modeling 
of the H {\sc ii} region. 
We summarize our findings in Section \ref{sec:CON}.

\section{OBSERVATIONS\label{sec:OBS}}

\subsection{Near-infrared observations\label{sec:NIR}}

NIR spectra of the BCDs II~Zw~40, Mrk~71, Mrk~930, Mrk~996 and SBS~0335$-$052E 
were obtained with the 3.5 m APO telescope in conjunction with the TripleSpec 
spectrograph on the 
night of 2009 November 25. TripleSpec \citep{W04} is a cross-dispersed NIR 
spectrograph that provides simultaneous continuous wavelength coverage from 
0.90 to 2.46 $\mu$m in five spectral orders during a single exposure. 
A 1\farcs1$\times$43\arcsec\ slit was used, resulting in a resolving power
of 3500. 

Three A0V standard stars, HD 16286, HD 23504, HD 212170 were observed for flux
calibration and correction for telluric absorption. Spectra of Ar 
comparison arcs were obtained on the same date 
for wavelength calibration. Since all observed targets
are smaller than the length of the slit, the nod-on-slit 
technique was used to acquire the sky spectrum. Objects were observed
by nodding between two positions A and B along the slit, 
following the ABBA sequence, 
and with an integration time of 200s or 300s at each position. 
The journal of the NIR observations is given in Table \ref{tab2}.
We show there the observation date, the exposure time, the extraction
aperture of the one-dimensional spectrum and the average airmass.

We carried out the reduction of the data according to the following procedures.
The two-dimensional spectra were first cleaned for cosmic ray hits 
using the IRAF\footnote{IRAF is 
the Image Reduction and Analysis Facility distributed by the 
National Optical Astronomy Observatory, which is operated by the 
Association of Universities for Research in Astronomy (AURA) under 
cooperative agreement with the National Science Foundation (NSF).} 
routine CRMEDIAN. 
Then all A and B frames were separately coadded and the resulting B frame
was subtracted from the resulting A frame. Finally, the (negative) 
spectrum at position B was adjusted to the (positive) spectrum at 
position A and subtracted from it.
The same reduction scheme was applied to the standard stars.

We then use the IRAF
routines IDENTIFY, REIDENTIFY, FITCOORD, TRANSFORM to 
perform wavelength calibration and correction for distortion and tilt for each 
frame. For all galaxies, a one-dimensional spectrum was extracted from the 
two-dimensional frames using the APALL IRAF routine, within a 
1\farcs1$\times$6\arcsec\ rectangular aperture so as to include the
brightest star-forming regions. The redshift of our objects range from 
0.00036 for Mrk 71 to 0.0183 for Mrk 930, so that 1$\arcsec$ 
corresponds to 7 pc for the nearest object and 350 pc for the 
furthest object, adopting a Hubble constant of 75 km s$^{-1}$ Mpc$^{-1}$. 
For the nearer objects, II Zw 40, Mrk 71 and Mrk 996, the slit 
contains a single stellar cluster which dominates in the line emission. 
Thus, in 
II Zw 40 the slit is centered on the compact super-star cluster (SSC) A  
 \citep{V08}, in Mrk 71 on the brightest 
compact cluster A \citep{TI05}, and in Mrk 996 on the central cluster 
\citep{T96}. All these stellar clusters are unresolved on {\sl HST/WFPC2} 
images.
For the furthest galaxies, SBS 0335--052E and Mrk 930, the slit is also 
centered on the brightest cluster, but contains several other 
stellar clusters. The {\sl HST} images of 
Mrk 930 and SBS 0335--052E are shown in \citet{M98} and \citet{T97}, 
respectively. For SBS 0335--052E,  the slit is oriented at 
a position angle of --30\arcdeg\ 
and includes not only SSCs 1 and 2, which dominate the 
line emission, but also SSCs 3--6. The spectrum of 
of Mrk 930 (its southern knot) was obtained 
at a position angle of +238\arcdeg. Because of the compact structure
of SBS 0335--052E and Mrk 930, we have not attempted to extract spectra of 
individual clusters and have used the integrated spectra. 
For these two furthest objects, the larger linear sizes of the slit also mean
that a larger area of the weakly ionized and 
neutral ISM around the stellar clusters is included 
in the slit, a fact which will be relevant in our modelization 
of the line emission of the low-ionization and neutral species 
(Section \ref{sec:CLO}) and in our discussion of the MIR extinction 
(Section \ref{sec:MIR}).    


Flux calibration and correction for telluric absorption were performed 
by first multiplying the 
one-dimensional spectrum of each galaxy by the synthetic absolute 
spectral distribution of the brightest standard star HD 23504, 
smoothed to the same
spectral resolution, and then by dividing the result by 
the observed one-dimensional spectrum of the same star.
Since there does not exist any published absolute spectral energy distribution
of standard stars, we have derived the one of HD 23504 by scaling   
the synthetic absolute spectral energy 
distribution of the star Vega ($\alpha$ Lyrae), of  
similar spectral type A0V, to the brightness of HD 23504.

\begin{figure*}
\figurenum{1}
\hbox{\includegraphics[angle=0,width=0.9\linewidth]{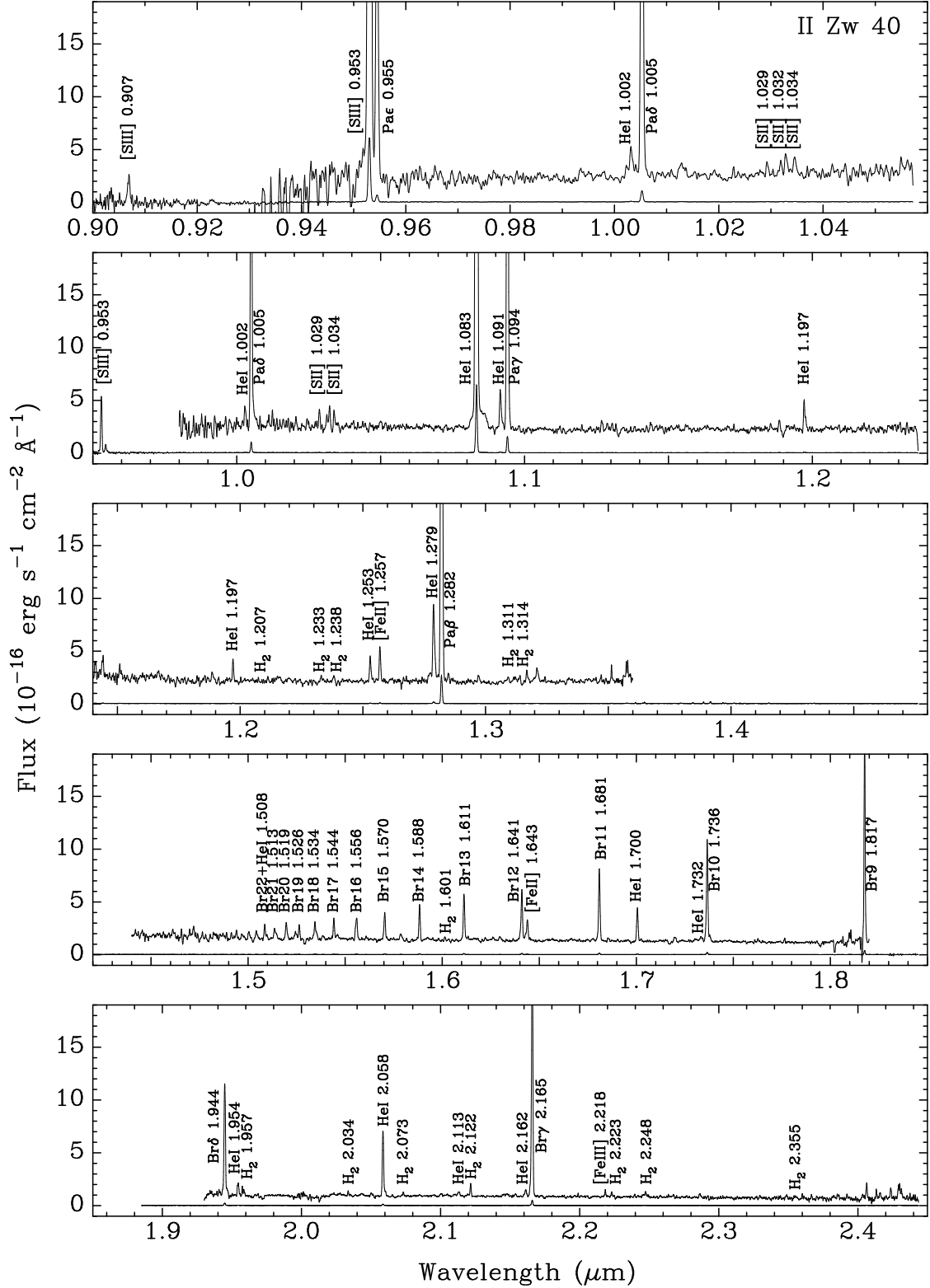} 
}
\figcaption{3.5m APO/TripleSpec NIR spectrum of II Zw 40 in five orders. 
In each panel, the noisy regions of the upper spectrum are omitted.  
They are caused by insufficient sensitivity or strong telluric
absorption. The flux scale on the y-axis corresponds 
to the upper spectrum. The lower spectrum is downscaled by a factor of 50
as compared to the upper spectrum. It is shown for the whole wavelength
range in each order.
\label{fig1}}
\end{figure*}

\begin{figure*}
\figurenum{2}
\hbox{\includegraphics[angle=0,width=0.9\linewidth]{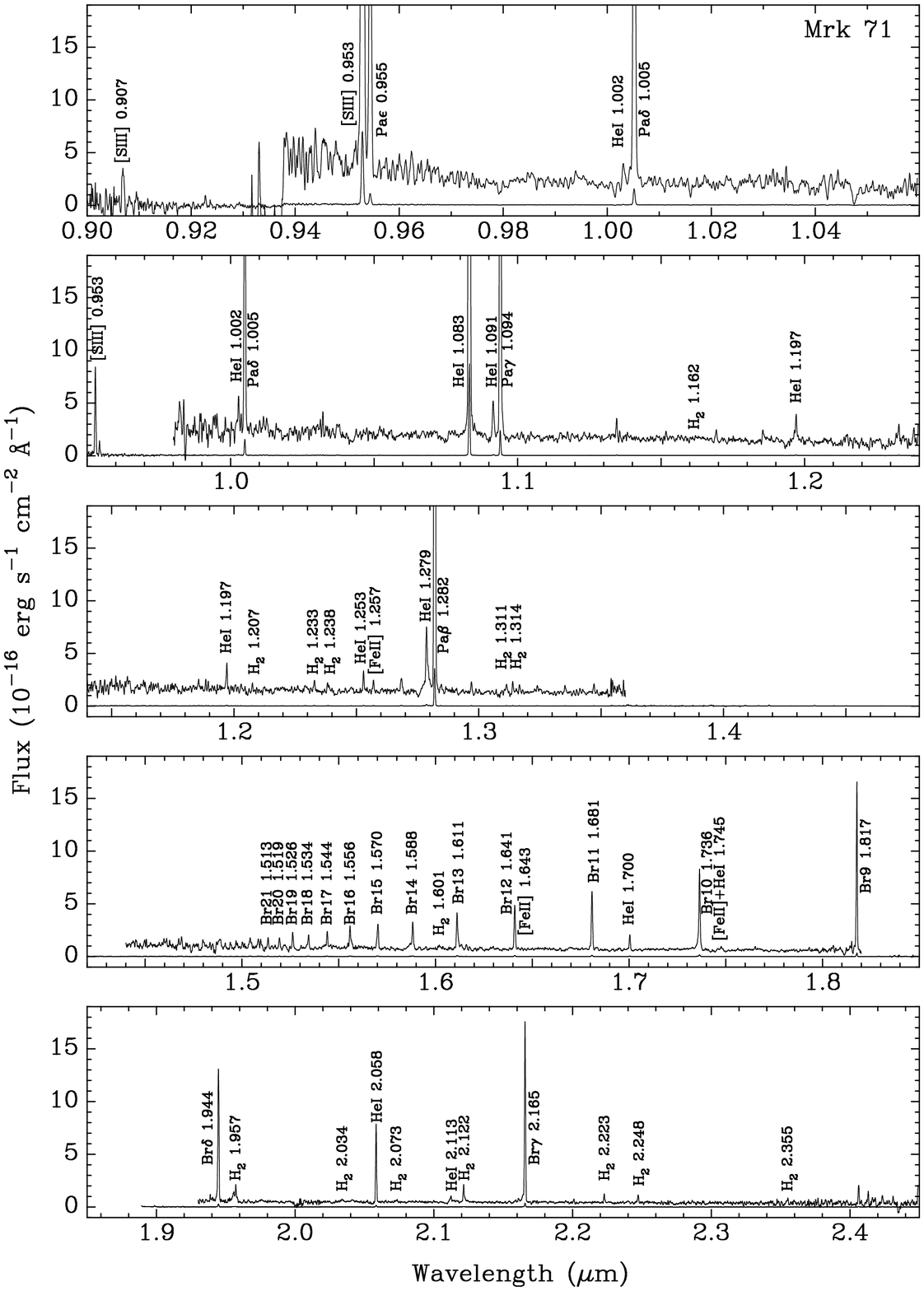} 
}
\figcaption{Same as in Fig. \ref{fig1}, but for Mrk 71.
\label{fig2}}
\end{figure*}

\begin{figure*}
\figurenum{3}
\hbox{\includegraphics[angle=0,width=0.9\linewidth]{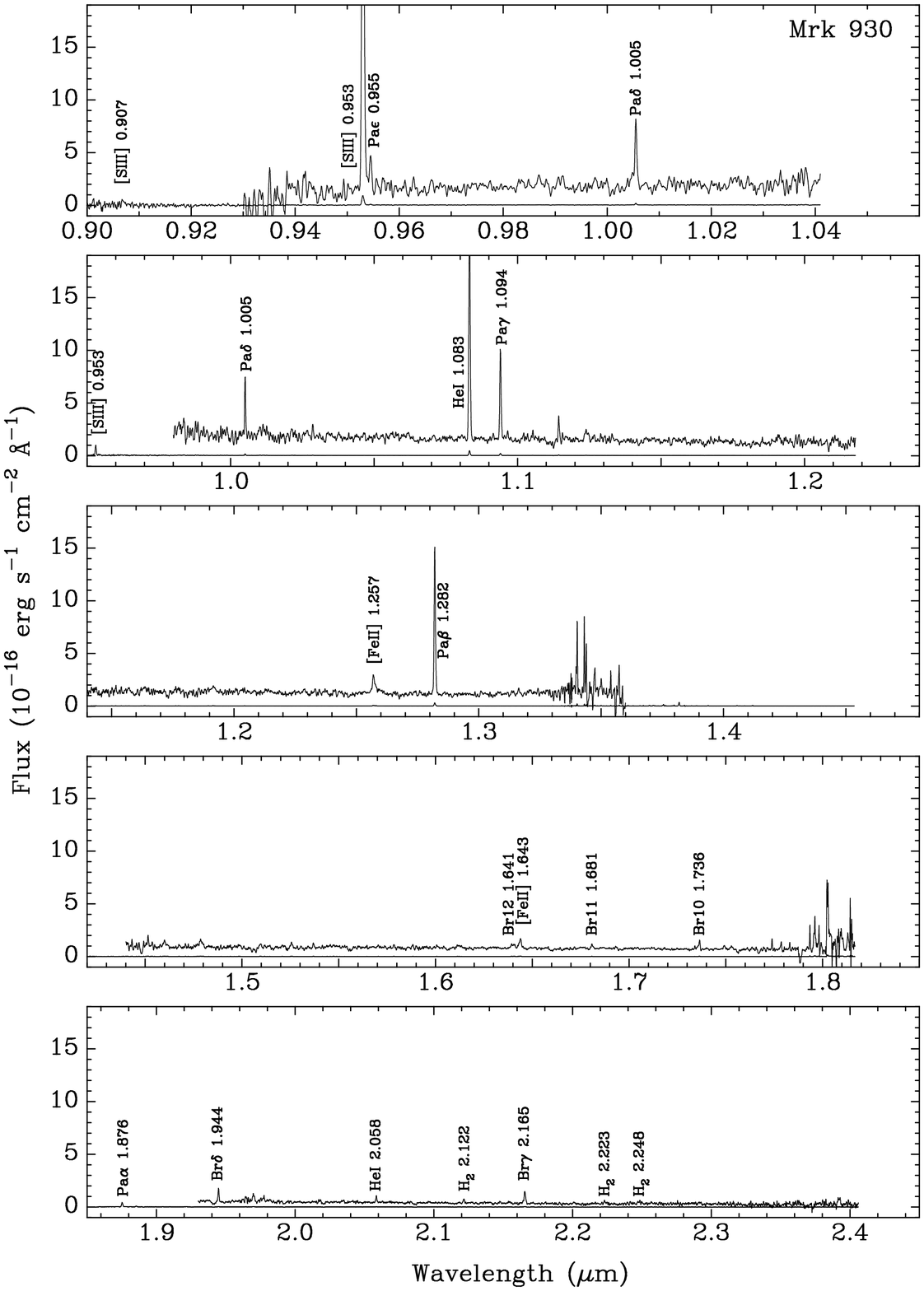} 
}
\figcaption{Same as in Fig. \ref{fig1}, but for Mrk 930.
\label{fig3}}
\end{figure*}

\begin{figure*}
\figurenum{4}
\hbox{\includegraphics[angle=0,width=0.9\linewidth]{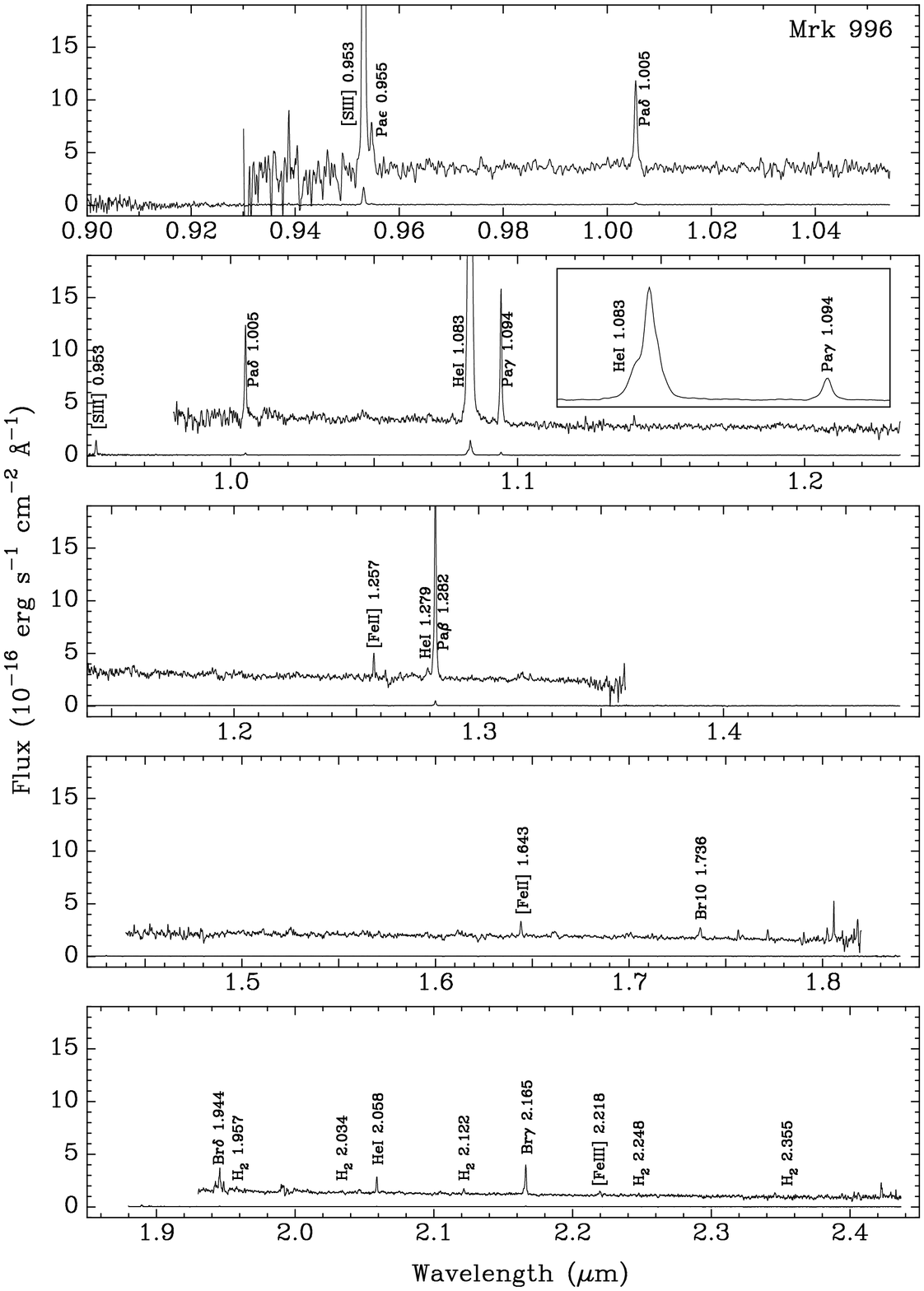} 
}
\figcaption{Same as in Fig. \ref{fig1}, but for Mrk 996.
\label{fig4}}
\end{figure*}

\begin{figure*}
\figurenum{5}
\hbox{\includegraphics[angle=0,width=0.9\linewidth]{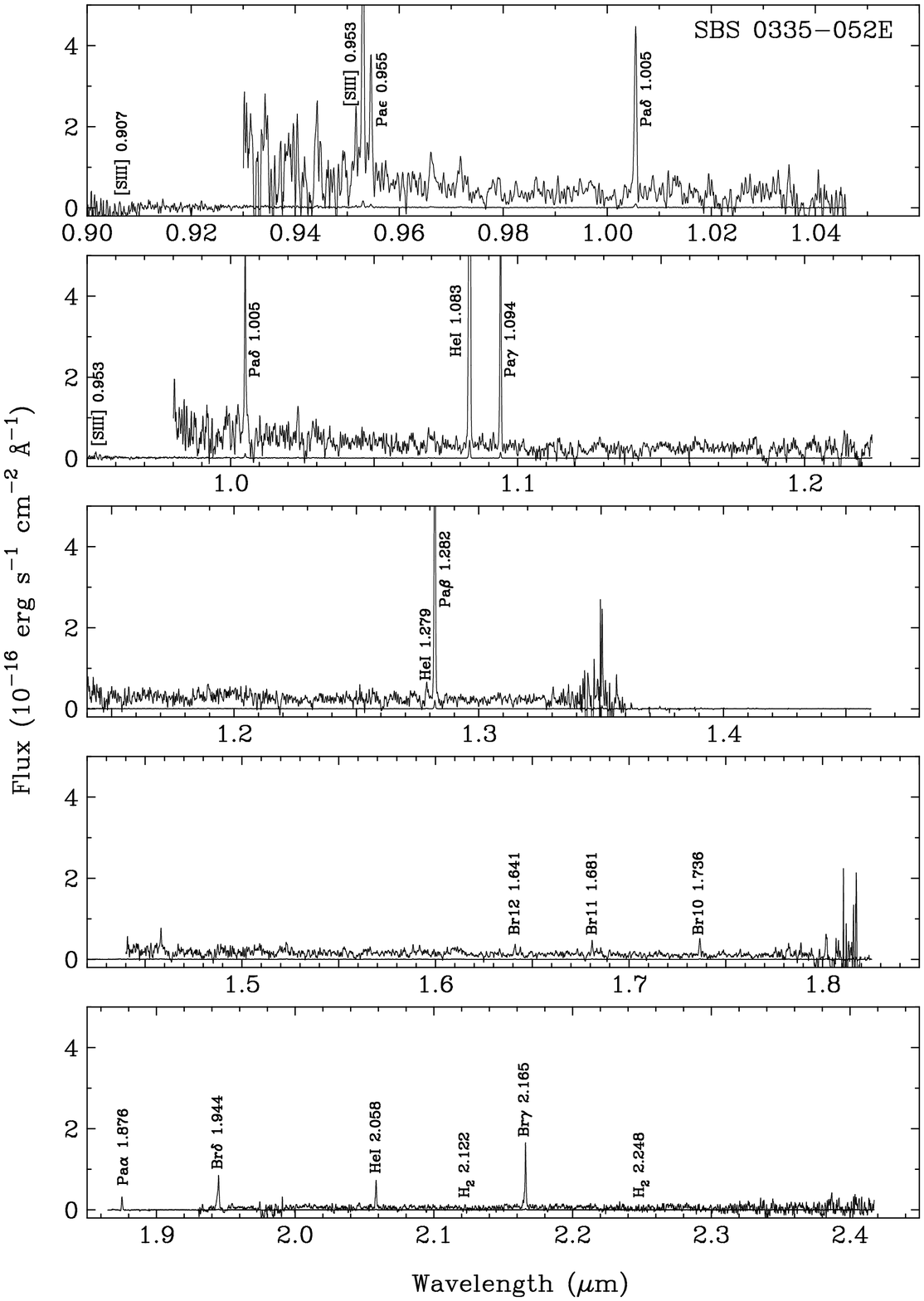} 
}
\figcaption{Same as in Fig. \ref{fig1}, but for SBS 0335$-$052E.
\label{fig5}}
\end{figure*}

The resulting 
one-dimensional flux-calibrated and redshift-corrected NIR spectra of all 
five BCDs are shown in Figs. \ref{fig1} -- \ref{fig5}.

\subsection{Optical observations\label{sec:OPT}}

To have a more complete physical picture of the extinction and 
of the ionization mechanisms in our five BCDs, we have supplemented the 
NIR observations with 
optical ones. The optical observations of all galaxies, with the exception 
of Mrk 930, 
were obtained with the 3.5 m APO telescope in the course of several nights
during the period November 2008 -- February 2010. 
For Mrk 930, we have used the optical spectrum obtained by \citet{IT98}. 
 
The 3.5 m APO observations were made with the Dual Imaging 
Spectrograph (DIS) in both the blue and red wavelength ranges. In the blue 
range, we use the B400 grating with a linear dispersion of 
1.83$\times$10$^{-4}$ $\mu$m/pix
and a central wavelength of 0.44 $\mu$m, while in the red range 
we use the R300 
grating with a linear dispersion 2.31$\times$10$^{-4}$ $\mu$m/pix and a 
central wavelength of 
0.75 $\mu$m. The above instrumental set-up gave a spatial scale along
the slit of 0\farcs4 pixel$^{-1}$, a spectral range $\sim$ 0.36 -- 0.97 $\mu$m 
and a spectral resolution of 7$\times$10$^{-4}$ $\mu$m (FWHM). Several Kitt 
Peak IRS spectroscopic 
standard stars were observed for flux calibration. Spectra of He-Ne-Ar 
comparison arcs were obtained at the beginning or the end of each night for 
wavelength calibration. The journal of the optical 
observations is given in 
Table \ref{tab2}. As for Mrk 930, the acquisition of its optical 
spectrum in the wavelength range
$\sim$ 0.36 -- 0.74 $\mu$m is described in \citet{IT98}.

The two-dimensional optical spectra were bias subtracted and flat-field 
corrected using IRAF. We then use the IRAF software routines IDENTIFY, 
REIDENTIFY, FITCOORD, TRANSFORM to perform wavelength calibration and 
correction  
for distortion and tilt for each frame. One-dimensional spectra were then 
extracted from each frame using the APALL routine. 
The extraction apertures for the one-dimensional optical spectra are shown in 
Table \ref{tab2}. The areas of these optical 
extraction apertures are generally similar
to those of the NIR extraction apertures. 
An exception is Mrk 930, for which the area
of the optical extraction aperture is $\sim$ 2 times that of 
the NIR extraction aperture.
The sensitivity curve was 
obtained by fitting with a high-order polynomial the observed spectral energy 
distribution of standard stars. Then sensitivity curves obtained
from observations of different stars during the same night were averaged.
The flux deviation from curve to curve is less than 1-2\%.

We have used the optical spectra to derive 
the physical conditions and oxygen abundances
of all five BCDs, following the procedures described
by \citet{I06}. The derived oxygen abundances are shown in Table \ref{tab1}.

The NIR and optical spectra of all BCDs, except for Mrk 930, overlap in
the  0.90 -- 0.97 $\mu$m wavelength range, allowing us to
match the fluxes of strong emission lines that are common in the two 
types of spectra. Specifically, we have adjusted 
the optical and NIR spectra so that 
the flux of the strongest line in the overlapping region, 
[S {\sc iii}] 0.953 $\mu$m, is 
the same in both spectra. 
We have thus multiplied the NIR spectra of II Zw 40, Mrk 71, Mrk 996 and
SBS 0335--052E by factors of 1.27, 1.17, 1.10 and 1.09, respectively. The 
multiplying factors are close to unity, as expected 
from the similarities of the areas of the optical and NIR 
extraction apertures. The deviations from unity can be caused 
by the slighly differing extraction apertures, non-perfect pointing and
varying weather conditions (transparency, seeing). They can also be caused by 
uncertainties in the flux calibration of the 
NIR spectra introduced by uncertainties in the absolute spectral 
energy distribution (SED) of the NIR standard star
HD 23504.
As discussed in Sect. \ref{sec:NIR}, this SED is not known and was obtained 
by scaling the synthethic NIR 
spectrum of Vega. 

Additional uncertainties are introduced by the ionization 
structure of the  H {\sc ii} regions, i.e. differences in the spatial 
distribution of various ions.
In particular, hydrogen line emission is more
concentrated to the center of the star-forming region than the [S {\sc iii}]
emission. We find that the FWHM of the H$\alpha$ emission
along the spatial axis of our objects is $\sim$20\% smaller than that
of the [S {\sc iii}] $\lambda$0.953$\mu$m emission. The ionization structure 
uncertainties 
are difficult to estimate as the 
observations were obtained with different slit widths,
1\farcs5 for optical spectra and 1\farcs1 for NIR ones.
We have attempted to minimize these uncertainties by  
extracting one-dimensional spectra
within apertures with similar areas.

\begin{figure*}
\figurenum{6}
\hbox{\includegraphics[angle=-90,width=1.0\linewidth]{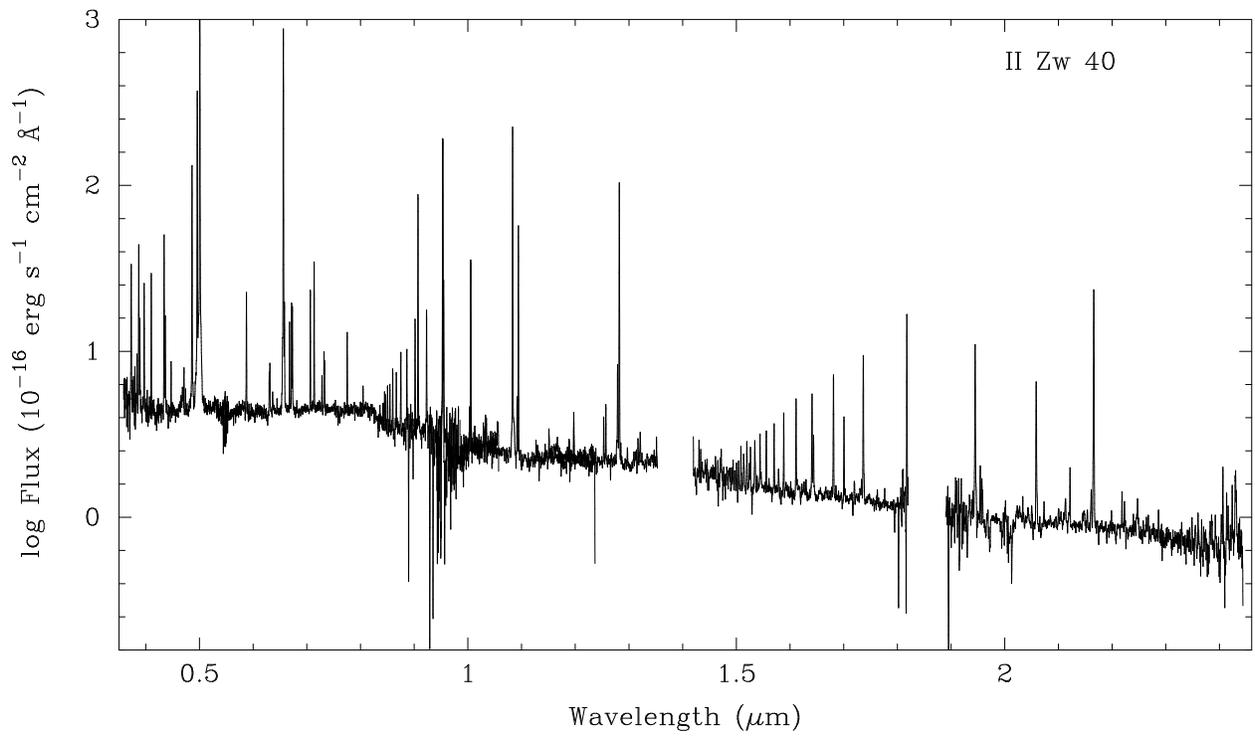}} 
\figcaption{Redshift-corrected spectral energy distribution of II Zw 40 in the 
0.36 $\mu$m -- 2.4 $\mu$m\ wavelength range \label{fig6}}
\end{figure*}

\begin{figure*}
\figurenum{7}
\hbox{\includegraphics[angle=-90,width=1.0\linewidth]{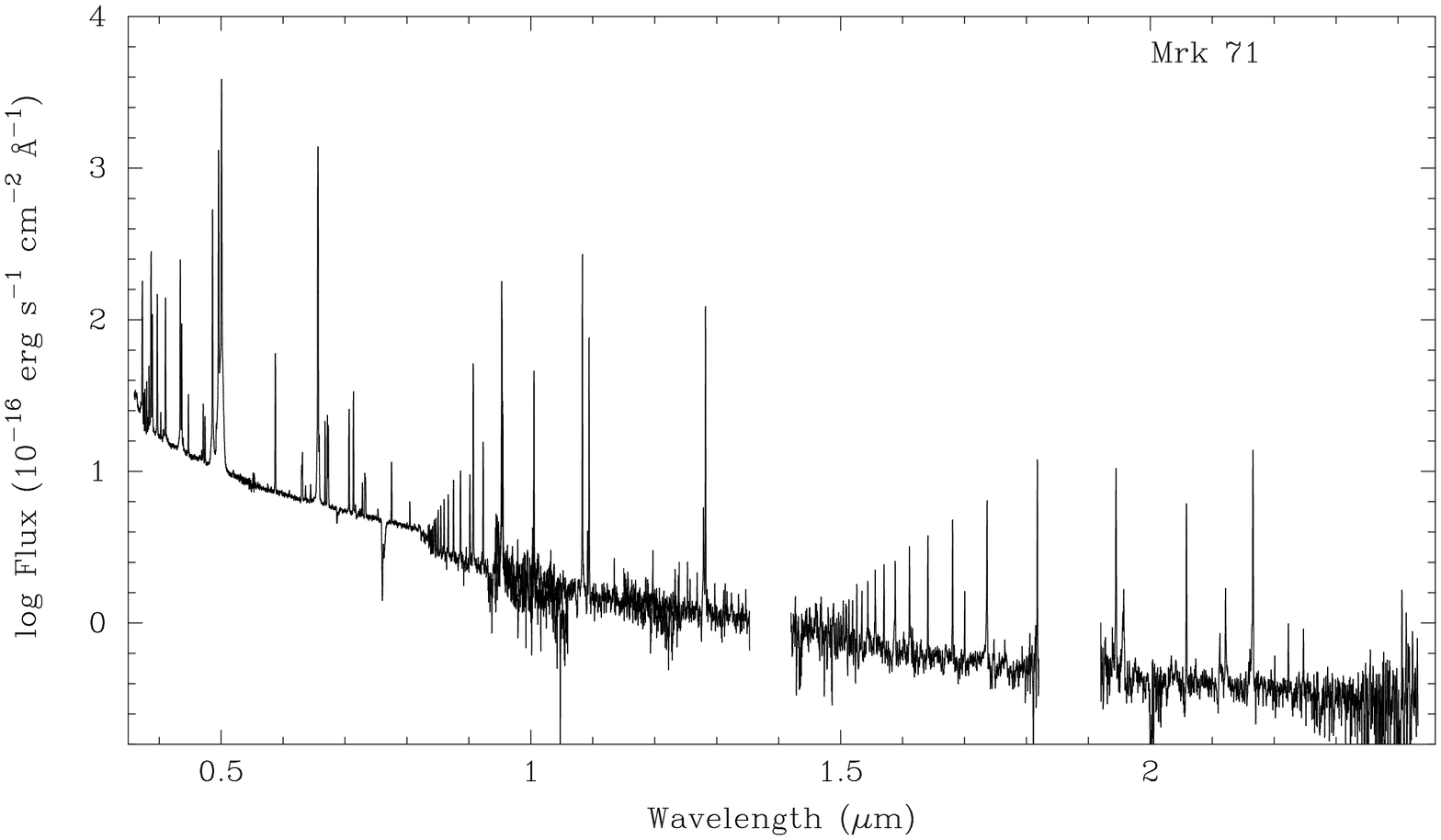}} 
\figcaption{Same as in Fig. \ref{fig6}, but for Mrk 71. \label{fig7}}
\end{figure*}

\begin{figure*}
\figurenum{8}
\hbox{\includegraphics[angle=-90,width=1.0\linewidth]{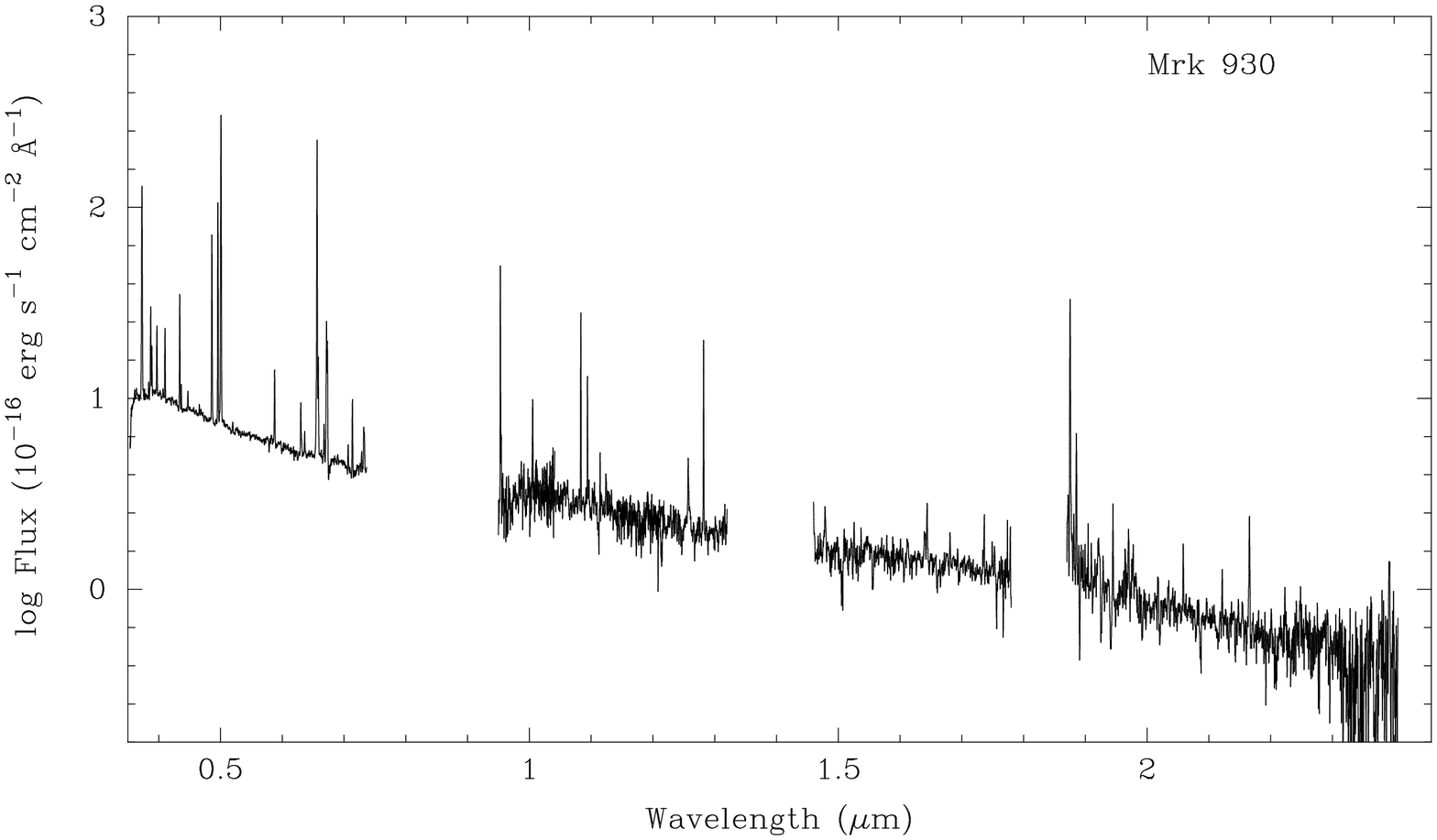}} 
\figcaption{Same as in Fig. \ref{fig6}, but for Mrk 930. \label{fig8}}
\end{figure*}

\begin{figure*}
\figurenum{9}
\hbox{\includegraphics[angle=-90,width=1.0\linewidth]{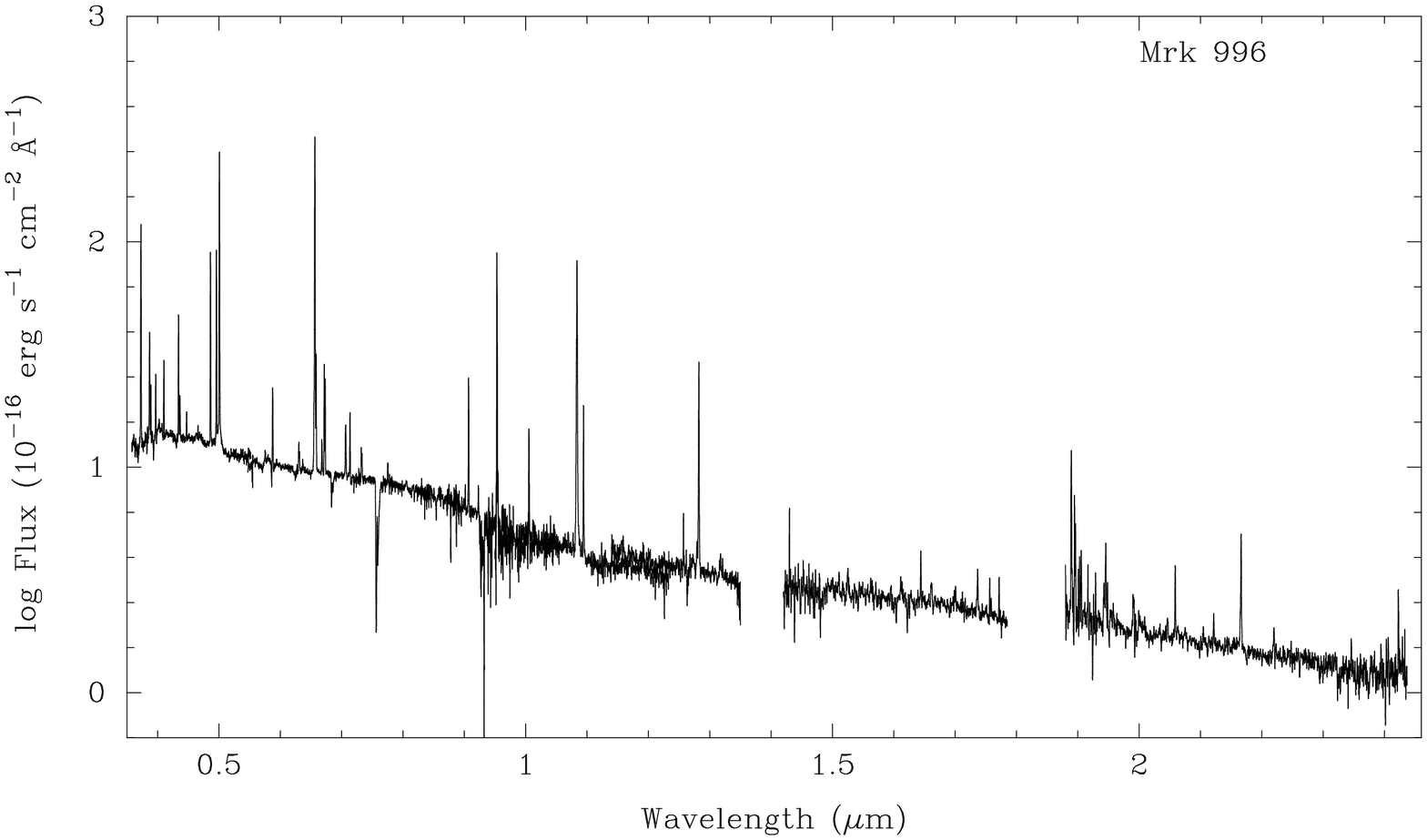}} 
\figcaption{Same as in Fig. \ref{fig6}, but for Mrk 996. \label{fig9}}
\end{figure*}

\begin{figure*}
\figurenum{10}
\hbox{\includegraphics[angle=-90,width=1.0\linewidth]{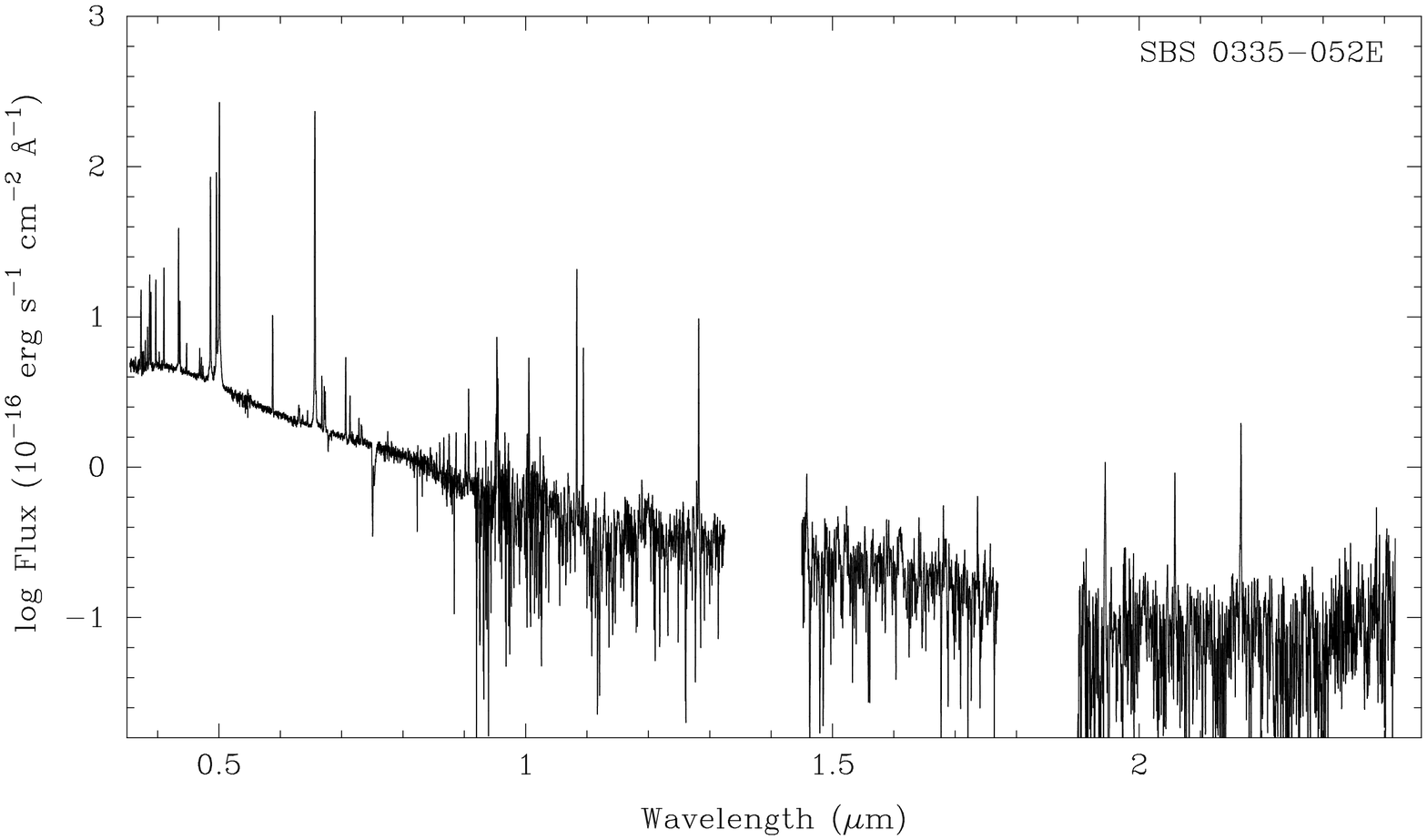}} 
\figcaption{Same as in Fig. \ref{fig6}, but for SBS 0335$-$052E. \label{fig10}}
\end{figure*}

For Mrk 930, we have scaled the NIR spectrum by a factor of 1.73 so that 
the Br$\gamma$/H$\beta$ flux ratio is equal to the theoretical recombination
value, after correcting the 
emission line fluxes for interstellar extinction
and underlying stellar absorption derived from the optical spectrum.
This scaling factor agrees nicely with the ratio of the areas 
of the apertures
used for the NIR and optical observations of Mrk 930.
The combined redshift-corrected
optical and NIR spectra for all BCDs 
are shown in Figs. \ref{fig6} -- \ref{fig10}. 
Gaps in the NIR spectra 
correspond to regions of strong absorption by telluric lines.
While the scaling of the NIR spectrum of Mrk 930 is more uncertain,  
it seems justified by the fact that NIR SED extends smoothly 
the optical SED, without any abrupt discontinuity (Fig. \ref{fig8}).

\section{RESULTS AND DISCUSSION\label{sec:RES}}

Emission-line fluxes in both the optical and NIR ranges were measured by using 
the SPLOT routine in IRAF. 
The errors of the line fluxes were calculated from the photon statistics
in the non-flux-calibrated spectra. They do not include systematic errors 
such as deviations of the standard star SED from the Vega SED, 
different apertures used
in the optical and NIR observations and varying weather
conditions. Those are difficult to estimate as discussed in 
Sect. \ref{sec:OPT}.
The observed fluxes $F$($\lambda$) of emission lines derived from the adjusted 
optical and NIR spectra 
are shown in the second columns of Tables \ref{tab3} -- \ref{tab7}. 
They are normalized to the observed H$\beta$ flux, $F$(H$\beta$),
shown at the end of each table, and multiplied by a factor of 100. 

How do our observations compare with previous ones of the 
same objects? II Zw 40 was observed by \citet{JL88} and 
\citet{C01} in the $K$ band, and 
by \citet{MO88}, \citet{V96} and \citet{V08} in the $H$ and $K$ bands. 
\citet{JL88} (3\farcs6 aperture), \citet{V96} (2\farcs4 aperture)
and \citet{C01} (2--3\arcsec\ aperture) obtained a flux of
the Br$\gamma$ emission line $F$(Br$\gamma$) in the 
range (3.4--3.9)$\times$10$^{-14}$
erg s$^{-1}$ cm$^{-2}$, slightly larger than our value of
3.1$\times$10$^{-14}$ erg s$^{-1}$ cm$^{-2}$ 
(1\farcs1$\times$6\arcsec\ aperture). On the other hand, \citet{MO88}
found $F$(Br$\gamma$) = 6.7$\times$10$^{-14}$ erg s$^{-1}$ cm$^{-2}$ in a 
6\arcsec$\times$6\arcsec\ aperture, while
\citet{V08} derived 
$F$(Br$\gamma$) = 1.0$\times$10$^{-14}$ erg s$^{-1}$ cm$^{-2}$ within a 
circular aperture of 1\arcsec\ in radius. SBS 0335--052E has been observed
in the NIR only by \citet{V00} who derived 
$F$(Br$\gamma$) = 1.5$\times$10$^{-15}$ erg s$^{-1}$ cm$^{-2}$ 
in an aperture of 1\arcsec$\times$1\farcs5, lower than our value of 
2.4$\times$10$^{-15}$ erg s$^{-1}$ cm$^{-2}$. All 
observations of II Zw 40 and SBS 0335--052E appear to be 
consistent in the sense
that the Br$\gamma$ fluxes obtained with apertures having similar areas 
are close 
to each other while those obtained with smaller apertures are lower.

The presence of many emission lines in the optical and NIR spectra
(Figs. \ref{fig1} -- \ref{fig10} and Tables \ref{tab3} -- \ref{tab7}) 
allows us to study  
extinction, H$_2$ and [Fe {\sc ii}] emission  
and excitation mechanisms in all five galaxies. We discuss these issues in the 
following sections.

\subsection{Optical and NIR extinction and hidden star formation\label{sec:EXT}}


In previous optical-NIR studies of BCDs, except for our own work on Mrk 59  
\citep{I09}, $JHK$ spectra have been observed separately, 
and there has been no wavelength overlap between the 
optical and NIR spectra. This introduces uncertainties, 
due to the use of generally different apertures in the optical and 
NIR observations, and
due to the adjusting of the continuum levels of 
the NIR spectra obtained separately in different orders. 
Because our NIR spectra have been obtained 
simultaneously over the whole $JHK$ wavelength  
range and because it is possible to directly match the optical and NIR 
spectra using common emission line fluxes (except for Mrk 930), 
these adjusting uncertainties
are mostly eliminated. This allows 
us to directly compare the optical and NIR extinctions and search for any 
hidden extinction at longer wavelengths.  
 
In the second column of Tables \ref{tab3} -- \ref{tab7}, we list the observed 
fluxes $F$($\lambda$) 
of the emission lines in both the optical and NIR spectra. In the third column
of the same Tables are shown the fluxes $I$($\lambda$), corrected for
underlying stellar absorption (for hydrogen lines) and extinction (for all
lines). The extinction coefficient $C$(H$\beta$) and equivalent width
EW(abs) of hydrogen lines derived from the decrement of hydrogen Balmer
lines in the optical spectrum are listed in the footnotes to  
Tables \ref{tab3} -- \ref{tab7}. They are
used to correct the emission line fluxes in both the optical
and NIR ranges, adopting the extinction curve of \citet{C89} and 
$R(V)$ = $A(V)/E(B-V)$ = 3.2, where $A(V)$ and $E(B-V)$ are respectively 
the $V$ band extinction
and the $B-V$ reddening. The value $R(V)$ =3.2 is 
typical for the diffuse interstellar medium in the Galaxy. With this value of 
$R(V)$, then $C$(H$\beta$) = 1.47$\times$$E(B-V)$ = 0.46$\times$$A(V)$.

However, the dust composition in the low-metallicity interstellar medium
in our BCDs 
can be different from that in the Galaxy, resulting in a different extinction 
curve. Indeed, \citet{R81} found that the UV extinction curve in the 
Small Magellanic Cloud (SMC) is significantly steeper than that in the Galaxy, 
indicating the presence of a larger fraction of small dust grains in the 
SMC. To reproduce the SMC extinction curve in the UV range, the value of $R(V)$
should be smaller than 3.2. However, differences
in the extinction curves with different $R(V)$s are much smaller 
in the optical and NIR ranges. 
For example, if we adopt $R(V)$ = 2.5, corresponding to  
$C$(H$\beta$) = 1.17$\times$$E(B-V)$ = 0.47$\times$$A(V)$, then 
$C$(H$\beta$) is decreased by $\sim$ 10\% and the flux of the 
Br$\gamma$ emission line relative to H$\beta$ is increased only by 
$\sim$ 2 -- 3\%, much smaller than the errors given in 
Tables \ref{tab3} -- \ref{tab7} for that line. Therefore, in the following,
we will use the extinction-corrected emission-line fluxes obtained with
$R(V)$ = 3.2.

The corrected fluxes of hydrogen lines in the third column of 
the same Tables are derived, assuming that the
difference between the observed and theoretical hydrogen Balmer
decrements is solely due to extinction and underlying stellar absorption.
However, collisional excitation of hydrogen lines may play a role, resulting
in a steepening of the 
observed Balmer decrement. Following \citet{L09}, we assume
that the fraction of the H$\alpha$ emission due to collisional
excitation is 5\%, while that for H$\beta$,
H$\gamma$ and H$\beta$ emission is 3\%. These values should be considered 
as upper limits for collisional excitation. A possible exception
is Mrk 996 with a very high-density H {\sc ii} region \citep{T08}, where 
collisional excitation could be higher. Subtracting the 
collisional contribution from the H$\alpha$, H$\beta$, H$\gamma$ and
H$\delta$ emission lines and correcting all lines for
extinction and hydrogen lines for underlying stellar absorption, 
we obtain the emission line
fluxes listed in the fourth column of Tables \ref{tab3} -- \ref{tab7}.

In the second to sixth columns of Table \ref{tab8}, we show the 
extinction-corrected fluxes $I$($\lambda$) 
of the strongest hydrogen lines in the optical and NIR spectra
for all observed galaxies. 
For comparison, the seventh column of Table \ref{tab8} shows the theoretical 
recombination flux ratios, calculated by \citet{HS87} for case B, with
an electron temperature $T_e$ = 15000 K and an electron number density
$N_e$ = 100 cm$^{-3}$. It can be seen that 
there is a very good agreement between
the corrected and theoretical recombination values of the optical and 
NIR line fluxes. 
We have shown in Table \ref{tab9} 
the extinction coefficients $C$(H$\beta$) derived from the ratio of the 
flux of each individual
hydrogen line to the H$\beta$ flux. They were derived 
from the equation
\begin{equation}
\frac{I(\lambda)}{I({\rm H}\beta)}=\frac{F(\lambda)}{F({\rm H}\beta)}
\frac{EW(\lambda)+EW(abs)}{EW(\lambda)}
\frac{EW({\rm H}\beta)}{EW({\rm H}\beta)+EW(abs)}
\times10^{f(\lambda)C({\rm H}\beta)}, \label{eq:chb}
\end{equation}
where $f(\lambda)$ is the value of the normalized reddening curve from \citet{C89}
at the wavelength $\lambda$, adopting $A(V)/E(B-V)$ = 3.2, 
$F$ and $I$ are respectively the 
observed and theoretical fluxes, EW($\lambda$) and
EW(H$\beta$) are equivalent widths of emission lines, and 
EW(abs) is the equivalent
width of the absorption hydrogen lines.
 We adopt EW(abs) = 2.7\AA, corresponding
to a starburst age of $\leq$ 4 Myr \citep{GD99}.

Inspection of Table \ref{tab9} shows that, for each object, 
the extinction coefficients
$C$(H$\beta$) derived from different hydrogen lines agree within the errors 
and do not show any systematic trend. This implies that the same  
extinction coefficient $C$(H$\beta$) [or extinction $A(V)$] holds over 
the whole 0.36 -- 2.40 $\mu$m wavelength range. 
The fact that $A(V)$ does not increase at longer 
wavelengths implies that there is no hidden star formation in the NIR 
as compared to in the optical range. 
This appears to be a general result for BCDs 
\citep[e.g.][]{V00,V02,I09}. 
This is true even in the extreme case of the BCD SBS 0335--052E 
where observations of the 
far-infrared (FIR) continuum imply a very large extinction 
[$A(V)$ $\sim$ 15--20 mag] and three times as much hidden star formation 
in the MIR than in the visible \citep{T99,H04}. We 
will explore the MIR extinction in section \ref{sec:MIR}.

\subsection{H$_2$ emission\label{sec:H2}}

Molecular hydrogen lines do not originate in the H {\sc ii} region, but in 
neutral molecular clouds. In the NIR, they are excited by two
mechanisms. The first one is the thermal mechanism consisting of 
collisions between neutral species (e.g., H, H$_2$), 
resulting from large-scale shocks 
driven by powerful stellar winds, supernova remnants 
or molecular cloud collisions. The second one is 
the fluorescent mechanism due to absorption of ultraviolet photons.
It is known that these two mechanisms excite mostly different
roto-vibrational levels of H$_2$. By comparing the observed line ratios with 
those predicted by models such as those calculated by \citet{BD87}, it is 
possible to discriminate between the two processes. 
In particular, line emission from the vibrational level $v$=2 and higher 
vibrational levels are virtually absent in 
collisionally excited spectra, while they are relatively strong in fluorescent
spectra.

H$_2$ emission lines are detected in the NIR spectra of all our objects
(see Figs. 1--5). 
Only one H$_2$ emission line, 2.122 $\mu$m 1-0 S(1), is definitely
detected in the spectrum of SBS 0335--052E. 
Another line, 2.248 $\mu$m 2-1 S(1), is  marginally
detected, although it is clearly present in the spectrum
of \citet{V00}.
Three, five, thirteen and fourteen H$_2$ lines are respectively 
detected in the NIR spectra of Mrk 930, Mrk 996, 
II Zw 40 and Mrk 71.
The fluxes of the H$_2$ lines relative to that of the strongest
2.122 $\mu$m 1-0 S(1) line are shown in Table \ref{tab10}. In the 
last two columns are shown the theoretical ratios calculated by \citet{BD87} 
in the cases of fluorescent and collisional excitation.
It is seen that the observed line flux ratios are
in agreement with those predicted for fluorescent excitation,
despite the rather high errors in the flux ratios.
Our finding is 
in agreement with the conclusions of \citet{V00} for SBS 0335--052E,
of \citet{V08} for II Zw 40 and of 
\citet{I09} for Mrk 59, where it was also found
that the fluorescence process is the main excitation mechanism   
of H$_2$ lines. This conclusion appears 
to be common for BCDs with high-excitation spectra.

\subsection{CLOUDY stellar photoionization modeling of the H {\sc ii} region
\label{sec:CLO}}

We next examine the excitation mechanisms of the NIR emission lines arising in
the H {\sc ii} regions of our BCDs. For this purpose, we have constructed 
stellar photoionization models for the H {\sc ii} regions using
the CLOUDY code (version 07.02.01) of \citet{F96} and \citet{F98}.  
We list in Table \ref{tab11} the input parameters of these models. 
The first row gives $Q$(H), the logarithm of
the number of ionizing photons per second, 
calculated from the H$\beta$ luminosity with the distance given in 
Table \ref{tab1}. For the determination of the H$\beta$
luminosity, we have taken 
the H$\beta$ fluxes from large aperture spectroscopic and photometric 
observations. We thus 
adopt the H$\beta$ fluxes to be 
3.5$\times$10$^{-13}$ erg s$^{-1}$ cm$^{-2}$ for II Zw 40 \citep{L07},
2.1$\times$10$^{-12}$ erg s$^{-1}$ cm$^{-2}$ for Mrk 71 \citep{MK06} and 
2.5$\times$10$^{-13}$ erg s$^{-1}$ cm$^{-2}$ for Mrk 930 \citep{MK06}. 
These fluxes were then corrected for extinction, as 
derived from the spectroscopic observations. We also adopt
the extinction-corrected H$\beta$ fluxes of  
6.6$\times$10$^{-13}$ erg s$^{-1}$ cm$^{-2}$ for Mrk 996 \citep{T08}
and of 1.8$\times$10$^{-13}$ erg s$^{-1}$ cm$^{-2}$ for
SBS 0335--052E \citep{I06b}.

The other rows of Table \ref{tab11} list $T_{eff}$, the effective temperature
of stellar radiation, and $N_e$, the electron number density, assumed to be
constant with radius for all objects, except for Mrk 996, 
for which we have adopted the 
number density radial distribution of \citet{T08}. 
The fourth row gives $f$, the filling factor. 
The remaining input parameters are the number ratios of
different species to hydrogen, derived mainly from the spectroscopic
observations in this paper, except for the carbon
and silicon abundances. There are no strong emission lines
of these last two 
species in the optical and NIR ranges. We have therefore calculated 
their abundances, using the mean relations 
of C/O and Si/O vs. oxygen abundance derived 
for low-metallicity H {\sc ii} regions by \citet{G95a,G95b}.
For the ionizing stellar radiation, 
we have adopted a \citet{K79} stellar atmosphere model with $T_{eff}$=50000 K. 
We have also experimented with other forms of the stellar ionizing 
radiation such as those given by Starburst99 models \citep{L99} or by 
models based on the Geneva stellar evolutionary tracks with heavy element
mass fractions $Z$ = 0.001 -- 0.004 \citep{M94}. 
In all cases, we were able to reproduce well most of the observed line 
intensities by varying only slightly the input parameters. Thus 
our results are robust and do not depend sensitively on a particular  
adopted grid of models.

Most of the input parameters in Table \ref{tab11}, e.g. $Q$(H) and
chemical composition, are constrained by observations and are thus kept
unchanged. This is not the case for the filling factor $f$ which is 
therefore varied until the best agreement is found between 
the observed and predicted line intensities. 
The CLOUDY-predicted fluxes of the  
emission lines are shown in the fifth column of Tables \ref{tab3} -- 
\ref{tab7}. Comparison of the extinction-corrected observed
and predicted emission-line fluxes in both the optical and NIR ranges shows 
that, in general, the agreement between the observations and the CLOUDY 
predictions is satisfactory. This is especially true 
for the strong hydrogen, neutral 
helium and forbidden lines of doubly ionized species.
The agreement however is not as good for emission lines of singly ionized 
and neutral species which arise in the outer parts of H {\sc ii} regions or 
even beyond the ionized hydrogen regions (e.g. the 
[S {\sc ii}] emission
lines). For example, the line intensities of the [S {\sc ii}]$\lambda$$\lambda$
6717,6731 emission lines are underpredicted by a factor of 2 for the nearer BCDs (Mrk 71 and 
II Zw 40) and by a factor of 3--4 for the more distant ones (SBS 0335--052 and 
Mrk 930). The larger disagreement factor for the more distant BCDs is 
probably due to the fact that their extraction apertures are larger in 
linear size and include more of the weakly ionized and  
neutral ISM than the apertures of the nearer BCDs.
   
Some part of the disagreement can be caused by our too simplistic
modeling with the CLOUDY code. 
For all galaxies but Mrk 996, we have adopted a 
spherically symmetric H {\sc ii} region model with a single compact central 
ionizing source and an uniform density distribution.  
In reality, our galaxies 
are likely to contain several non-symmetric and non-uniform H {\sc ii} regions
with different ionization parameters which contribute in different proportions
to the excitation of high- and low-ionization species in the
integrated spectra. With the exception of Mrk 996, we do not have 
enough observational data to produce more 
complex and refined models of the H {\sc ii} regions. 
For the sole case of Mrk 996, we have adopted a non-uniform radial density 
distribution \citep[the one dervived by ][]{T96,T08}, and there is indeed 
a much better agreement between the observed and predicted line intensities, 
particularly for the [S {\sc ii}]$\lambda$$\lambda$
6717,6731 emission lines. 

More sophisticated modeling will certainly improve the overall agreement 
between observations and models. However, these improvements
will not change the main conclusion 
that H {\sc ii} region models including only stellar 
photoionization as an ionizing source can  
account for most of the observed line fluxes, 
both in the optical and NIR ranges.
For the great majority of the lines, no additional excitation mechanism 
such as shocks from stellar winds and supernova remnants 
is needed. This situation is again similar to that   
in the BCDs II Zw 40 and Mrk 59, where 
\citet{V08} and \citet{I09} found that the ISM is mainly 
photoexcited by stars. However, we note that while some He {\sc i} 
emission lines in the spectrum of Mrk 996 are reproduced reasonably well, the
modeled fluxes of the He {\sc i} $\lambda$0.707$\mu$m and 1.083$\mu$m emission 
lines are several times larger than those observed. These two lines are very 
sensitive to collisional and fluorescent excitation. In the very high-density
H {\sc ii} region of Mrk 996 \citep{T08}, 
the He {\sc i} $\lambda$1.083$\mu$m is optically thick. This may be a reason
why this line has an asymmetric profile (see inset in Fig. \ref{fig4}).
Therefore, a detailed treatment of the radiative
transfer in this line and of the collisional excitation of He {\sc i}
states is needed. 

\subsection{The contribution of shock photoionization \label{sec:shock}}

There are other exceptions. 
One such exception is the He {\sc ii} $\lambda$0.469 $\mu$m
nebular emission line. It
is detected in the optical spectra of three BCDs,  
Mrk 71, Mrk 930 and 
SBS 0335--052E, but cannot be reproduced by the photoionized H {\sc ii}
region models with only stellar radiation as the ionizing source.
In the two remaining BCDs, II Zw 40 and Mrk 996, this emission line is 
also detected, but it is broad, implying that it is produced by  
expanding envelopes
of Wolf-Rayet stars, and not by nebular emission.  To reproduce the
observed nebular He {\sc ii} $\lambda$0.469 $\mu$m 
emission-line flux, a harder ionizing radiation
is required. Therefore, for all BCDs (except for II Zw 40 and Mrk 996), 
we have also calculated models of H {\sc ii} regions with
a constant number density, which include ionization radiation
by fast shocks, in addition to the stellar ionizing radiation.  
We vary  
the number of ionizing photons produced by fast shocks until the 
model prediction best matches the observed flux of the nebular 
He {\sc ii} $\lambda$0.469 $\mu$m emission line. 
We adopt the spectral distribution of the
ionizing radiation to be that of a shock with a velocity of 500 km s$^{-1}$
\citep{A08}.
For best agreement, we found the number of ionizing photons from 
fast shocks to be respectively 2.5\%, 4\% and 13\% the number of ionizing
stellar photons for Mrk 71, Mrk 930 and SBS 0335--052E. 
 In the last column of Tables \ref{tab4}, \ref{tab5} 
and \ref{tab7}, we 
show the emission-line fluxes predicted by combined models including
both stellar and shock ionizing radiation. These
models reproduce well 
the observed nebular He {\sc ii} $\lambda$0.469 $\mu$m 
emission line fluxes. They also increase the relative
fluxes of the low-ionization species, such as [O {\sc i}] $\lambda$0.630 and
[S {\sc ii}] $\lambda$0.672, 0.673 $\mu$m emission lines, making them 
agree better with the observations.  
On the other hand, the relative line  
fluxes of the high-ionization species such as [Ne {\sc iii}] $\lambda$0.387, 
and [S {\sc iii}] $\lambda$0.9533 $\mu$m 
are little changed. This is probably because 
the shock-produced harder ionizing radiation penetrates more efficiently
into the outer layers of the H {\sc ii} region, as compared to the softer
stellar ionizing radiation. This results in more efficient heating of the gas
and excitation of the low-ionization species.

\subsection{[Fe {\sc ii}] line emission\label{sec:FeII}}

The [Fe {\sc ii}] $\lambda$1.257 and $\lambda$1.643 $\mu$m emission
lines have often been used as indicators of the importance of shock excitation 
relative to that of photoionization \citep{MO88,O90,O01}. 
All of our galaxies either show one of them or both.
In the Galaxy, the intensities of these lines are considerably 
smaller in regions where photoionization dominates than in those 
where shock excitation dominates, such as in SN remnants. For example,      
the ratio of the [Fe {\sc ii}] $\lambda$1.643 $\mu$m line to Br$\gamma$ is 
0.06 in the Orion nebula, while it is more than 30 in Galactic SN remnants 
\citep{MO88,O90}.

The [Fe {\sc ii}] $\lambda$1.257 $\mu$m and $\lambda$1.643 $\mu$m lines to 
Br$\gamma$ ratios have the very 
low values of $\sim$ 0.05 -- 0.1 in II Zw 40, Mrk 71 and SBS 0335--052E, 
similar to the value of 0.055 found by \citet{I09} in Mrk 59, and to 
the one in the Orion nebula.  However, 
in the two
other galaxies, Mrk 930 and Mrk 996, which have lower excitation H {\sc ii}
regions, the [Fe {\sc ii}]/Br$\gamma$ ratios are 
considerably higher, with values $\sim$ 0.5, but  
still much lower than the value of $>$30 in Galactic SN remnants.
Does this mean that 
SN shock excitation does not play a role in all our BCDs? 

Examination of Table \ref{tab12} shows that, even without any 
contribution of shock excitation from SNe, the stellar photoionization 
CLOUDY models overpredict the [Fe {\sc ii}] $\lambda$1.257 and 
$\lambda$1.643 $\mu$m line fluxes by factors of $\sim$1.5 -- 3. 
An exception is Mrk 930, where the extinction-corrected [Fe {\sc ii}]
line fluxes are a factor of $\sim$ 2 greater than the modeled ones.
If shock ionizing radiation is included (last column of 
Table \ref{tab12}), then the modeled [Fe {\sc ii}] emission
line fluxes are more comparable to the observed ones in Mrk 930. 
However, they are much higher than the observed fluxes
in other galaxies. It appears that, for H {\sc ii} 
regions in II Zw 40, Mrk 71, Mrk 996 and SBS 0335--052E, there is no need to
invoke any shock ionizing radiation. 
 However, shocks may play an important role in the excitation
of [Fe {\sc ii}] emission lines in Mrk 930.

The overprediction of the [Fe {\sc ii}] fluxes in the four BCDs other than
Mrk 930 may be due to aperture effects.
The small apertures used in the 
optical and NIR observations (the slit widths are respectively 
1\farcs5 and 1\farcs1) include only the inner 
parts of the H {\sc ii} region  
while singly ionized iron is mainly located in its outer parts.
This conclusion appears to be supported by the fact that the 
discrepancy factor is largest for the nearest BCD, Mrk 71, at a distance 
of 3.4 Mpc, while it is lower
for more distant galaxies, with distances larger 
than 10 Mpc. The aperture effect is the smallest for the most
distant BCD, Mrk 930.
\citet{I09} have proposed another explanation for the BCD Mrk 59, at a 
distance of 10.7 Mpc. They argued 
that the overprediction by a factor of $\sim$ 2 
of the [Fe {\sc ii}] 1.257 and 1.643 $\mu$m emission line intensities by the 
CLOUDY stellar ionization model for this BCD can be explained by 
the depletion of iron atoms onto dust grains \citep{O01,I06}.
This depletion process is likely to be most efficient in the outer zones of 
the H {\sc ii} region where the [Fe {\sc ii}] emission lines arise. 
These have less intense ionizing radiation as compared to the inner zones where
the ionizing radiation is more intense and where the [Fe {\sc iii}] optical 
emission lines arise. It is likely that both aperture effects and depletion
of iron onto dust play a role in accounting for the discrepancies between 
the modeled and observed [Fe {\sc ii}] fluxes.

\subsection{Comparison with mid-infrared observations: 
MIR extinction\label{sec:MIR}}

We have shown above that the observed relative line flux ratios 
in both the optical and NIR ranges in all our objects can be 
satisfactorily accounted for with the 
same extinction coefficient $C$(H$\beta$) [or the same extinction
$A$($V$)]. In other words,  
there is no more hidden star formation in the NIR range as 
compared to in the optical range in regions traced
by emission lines, i.e. in ionized gas regions.
A question then arises: would we see more hidden star formation
at longer wavelengths, in the mid-infrared (MIR) range?  

To investigate the extinction in the MIR range,  
we have used the published {\sl Spitzer} MIR emission line fluxes 
of \citet{W06} and \citet{O06} for II Zw 40, 
of \citet{O06} for Mrk 930,
of \citet{T08} for Mrk 996, and of \citet{H04} 
for SBS 0335--052E.
{\sl Spitzer} observations of Mrk 71 exist but have not yet been published.
Since the {\sl Spitzer} spectra were obtained within significantly larger
apertures as compared to the optical and NIR spectra, we have normalized
the MIR emission-line fluxes to the optical and NIR fluxes by 
multiplying them by the ratios of the extinction-corrected 
optical H$\beta$ fluxes listed in Tables 3--7 to those 
obtained from large aperture spectroscopic observations or from
photometric observations. Thus, for II Zw 40 and Mrk 930 we have adopted the 
H$\beta$ fluxes of \citet{L07} and \citet{MK06}, respectively. 
These 
fluxes were then corrected for extinction with the extinction coefficient
$C$(H$\beta$) derived in this paper from the optical spectra. 
For Mrk 996 and SBS 0335--052E, we have 
adopted respectively the 
extinction-corrected H$\beta$ fluxes of \citet{T08} and \citet{I06}.
The multiplicative factors are respectively  0.32, 0.27, 0.31, and 
0.50 for II Zw 40, Mrk 930, Mrk 996 and SBS 0335--052E.
 No extinction correction has been applied to the MIR data as 
it is presumably small in this wavelength range.

  The normalized MIR emission line fluxes are shown in Table \ref{tab13}
together with the flux predictions by the CLOUDY models discussed 
in Section \ref{sec:CLO} for the optical and NIR ranges. 
Comparison of the MIR observed fluxes
with those predicted by the pure stellar ionizing radiation model 
shows agreement to within a factor of 2. 
The disagreement is the worst for the  
low ionization lines 
[Ne {\sc ii}] $\lambda$12.81 $\mu$m in II Zw 40 and  
[Fe {\sc ii}] $\lambda$25.99 $\mu$m in II Zw 40 and Mrk 930, 
the fluxes of which are predicted 
to be considerably smaller than the observed ones. 
Furthermore, the presence of 
the high-ionization emission line [O {\sc iv}] $\lambda$25.89 $\mu$m 
in Mrk 996 cannot be accounted for by a pure stellar ionizing radiation model 
\citep{T08}. 
The high observed flux of [Fe {\sc ii}] $\lambda$25.99 $\mu$m in Mrk 930
cannot be explained by shock excitation, but it can
be understood if we note that neutral iron has an ionization potential
smaller than that of hydrogen. Therefore, singly ionized iron may 
exist in weakly ionized cool interstellar medium,
outside the H {\sc ii} regions, and 
contribute to the [Fe {\sc ii}] $\lambda$25.99 $\mu$m emission, a 
contribution which is not accounted for in CLOUDY models. 
As for the high-ionization [O {\sc iv}] $\lambda$25.89 $\mu$m emission line in 
Mrk 996, 
\citet{T08} found that a CLOUDY model where the fraction 
of ionizing photons from shocks 
is only 3\% fits well all the MIR emission-lines 
of Mrk 996, including [O {\sc iv}] $\lambda$25.89 $\mu$m.
We give the predictions of that model for Mrk 996 in Table \ref{tab13}.    
The agreement within a factor of 2 between the observed fluxes of 
the MIR emission-lines and the predicted ones of CLOUDY models, 
based on the extinction-corrected optical emission lines,
implies that there 
is not considerably more hidden star formation seen in the MIR range, 
as compared to in the optical and NIR ranges.
This implies that, in low-metallicity BCDs, 
the H {\sc ii} regions with emission lines 
detected in the MIR range are in general relatively transparent, so they are 
also seen in the optical and NIR ranges. 

At first glance, 
this conclusion appears to 
be in flagrant disagreement with the findings of 
\citet{T99} and \citet{H04} who inferred  
highly obscured regions in SBS 0335--052E, with an $A(V)$ $\sim$ 15 mag,  
from modeling of the MIR continuum emission produced by warm dust.
Based on these observations, \citet{T99} concluded that some 75\% of the star
formation in SBS 0335--052E is hidden.
These two apparently discrepant results can be reconciled if the region 
where resides the warm dust responsible for the MIR emission is different from 
the ionized gas region where the emission lines arise.     
Studies of the H {\sc ii} regions
in the Milky Way \citep[e.g., ][]{Le07} show a good correlation
between e.g. the [S {\sc iv}] $\lambda$10.51$\mu$m emission line flux and the
warm dust continuum, implying that the warm dust resides in the ionized
medium. However, the physical conditions in SBS 0335--052E are very different
from those in the Milky Way H {\sc ii} regions. 
In particular, star formation in
SBS 0335--052E occurs in several compact SSCs
containing several thousand low-metallicity O stars \citep[e.g., ][]
{T97, J09}, 
responsible 
for a very high UV flux in the H {\sc ii} regions around the SSCs. 
It is doubtful whether dust grains can survive in such an extreme
environment. Furthermore, if the warm dust is 
in the same region as the ionized gas, we would expect an excess of the MIR 
emission line fluxes as compared to the predicted ones. No such excess 
is found, at least for 
the [Ne {\sc iii}] $\lambda$15.55 $\mu$m emission line
(Table \ref{tab13}). The 
dust is likely located in the neutral gas that surrounds the compact 
H {\sc ii} regions and which does not contribute to the 
observed MIR emission lines. In other words, the MIR emission lines 
and continuum arise in different regions, 
the first ones in relatively transparent H {\sc ii} regions 
and the second one in highly extincted regions surrounding 
the H {\sc ii} regions.

\section{SUMMARY AND CONCLUSIONS\label{sec:CON}}

We present here near-infrared (NIR) spectroscopic observations
of five blue compact dwarf (BCD) galaxies, II Zw 40, Mrk 71, Mrk 930, Mrk 996
and SBS 0335--052E, obtained with the 3.5m APO telescope. 
The NIR data have been supplemented
by optical spectra obtained with the 3.5m APO and 
the 2.1m KPNO telescopes, resulting
in a total wavelength coverage of 0.36 -- 2.46 $\mu$m. The 
overlap in the 0.90 - 0.97 $\mu$m wavelength range 
allows to adjust the continuum levels of the NIR spectra to the optical 
ones by insuring that the flux of the [S {\sc iii}] 
$\lambda$0.953 $\mu$m emission line is the same in both types of spectra.

We have arrived at the following conclusions:

1. In all BCDs, the extinction coefficient $C$(H$\beta$) = 
0.46$\times$$A$($V$) derived in the optical 
is the same as the one derived in the NIR for the ionized gas 
regions traced by emission lines. 
The NIR emission lines do not probe more extinct regions with 
hidden star formation as compared to the optical emission lines.
Using Spitzer MIR data, we have 
also found that the extinction $A$($V$) inferred from MIR observations 
is similar to the one derived from optical and NIR observations.
Thus, the emission-line spectra of low-metallicity BCDs in the whole 
$\sim$ 0.36 -- 25 $\mu$m wavelength range originates in relatively transparent 
H {\sc ii} regions. The large extinction deduced from the 
MIR continuum emission in 
the BCD SBS 0335--052E probably arises from 
neutral regions surrounding the H {\sc ii} regions. 

2. We have detected molecular hydrogen emission lines in the NIR spectra of 
all BCDs. Comparison of the observed fluxes with modeled ones 
suggests that the main excitation mechanism of H$_2$ emission
in all BCDs is fluorescence. 

3. A CLOUDY model with a pure stellar ionizing radiation
reproduces well the observed fluxes of most emission
lines in both optical and NIR ranges. Shocks are likely present in the 
H {\sc ii} 
region, but play a minor role in the ionization, at the 
level of 10\% or less, except for
high-ionization emission lines, such as 
He {\sc ii} $\lambda$0.469 $\mu$m in the optical range and
[O {\sc iv}] $\lambda$25.89 $\mu$m in the MIR range. 

4. CLOUDY stellar and shock ionizing models show that   
the [Fe {\sc ii}] $\lambda$1.257 and $\lambda$1.643 $\mu$m emission lines, 
often used 
as SN shock indicators in low-excitation high-metallicity starburst galaxies, 
cannot play such a role in high-excitation low-metallicity 
H {\sc ii} regions, 
such as those in the BCDs II Zw 40, Mrk 71, Mrk 996 and SBS 0335--052E. 
However, we find that shocks could be important contributors to the emission
of [Fe {\sc ii}] lines in Mrk 930.

\acknowledgements

The 3.5 APO time was available thanks to a grant from the Frank
Levinson Fund of the Silicon Valley Community Foundation to
the Astronomy Department of the University of Virginia.
Y.I.I. is grateful to the staff of the Astronomy Department at the 
University of Virginia for their warm hospitality. 
T.X.T acknowledges the support of NASA.




\begin{deluxetable}{lccccc}
  \tablecolumns{6}
\tabletypesize{\small}
  \tablewidth{0pt}
  \tablecaption{General characteristics of galaxies \label{tab1}}
  \tablehead{
\colhead{Object} & \colhead{R.A. (J2000.0)} & \colhead{Dec. (J2000.0)} 
& \colhead{Redshift}& \colhead{Distance\tablenotemark{a}}
& \colhead{12+logO/H\tablenotemark{b}}}
  \startdata
II Zw 40       & 05 55 42.6 & $+$03 23 32& 0.00263& 10.4 & 8.00 \\
Mrk 71         & 07 28 42.8 & $+$69 11 21& 0.00036&  3.4 & 7.85 \\
Mrk 930        & 23 31 58.3 & $+$28 56 50& 0.01830& 77.4 & 8.08 \\
Mrk 996        & 01 27 35.5 & $-$06 19 36& 0.00541& 21.8 & 8.10 \\
SBS 0335$-$052E& 03 37 44.0 & $-$05 02 40& 0.01348& 53.8 & 7.29 \\
  \enddata
\tablenotetext{a}{Distances of all galaxies but Mrk 71 are obtained from
the radial velocities corrected for the Virgo infall and are taken
from the NASA/IPAC Extragalactic Database (NED). 
The distance of Mrk 71 is from \citet{TI05}.}
\tablenotetext{b}{this paper.}
  \end{deluxetable}


\begin{deluxetable}{lrccc}
  \tablecolumns{4}
  \tablewidth{0pt}
\tabletypesize{\small}
  \tablecaption{Journal of observations \label{tab2}}
  \tablehead{
\colhead{Object} & \colhead{Date} & \colhead{Exposure (s)} & \colhead{Extraction}& \colhead{Airmass}\\
\colhead{      } & \colhead{    } & \colhead{            } & \colhead{aperture\tablenotemark{a}  }& \colhead{       }}
  \startdata
\multicolumn{5}{c}{Near-infrared observations} \\
II Zw 40       & 25 Nov. 2009 & 2000 (4$\times$300 + 4$\times$200)&1\farcs1$\times$6\arcsec\ &  1.7 \\
Mrk 71         & 25 Nov. 2009 & 1200 (4$\times$300)               &1\farcs1$\times$6\arcsec\ &  1.8 \\
Mrk 930        & 25 Nov. 2009 & 1200 (4$\times$300)               &1\farcs1$\times$6\arcsec\ &  1.0 \\
Mrk 996        & 25 Nov. 2009 & 2400 (8$\times$300)               &1\farcs1$\times$6\arcsec\ &  1.7 \\
SBS 0335$-$052E& 25 Nov. 2009 & 3600 (12$\times$300)              &1\farcs1$\times$6\arcsec\ &  1.7 \\
\multicolumn{5}{c}{Optical observations} \\
II Zw 40       & 07 Feb. 2010 & 1200                              &1\farcs5$\times$4\arcsec\ &  1.2 \\
Mrk 71         & 20 Nov. 2009 &  660                              &1\farcs5$\times$4\arcsec\ &  1.3 \\
Mrk 930        & 16 Dec. 1996 & 3600                              &2\arcsec$\times$6\farcs9  &  1.0 \\
Mrk 996        & 06 Nov. 2008 & 1200                              &1\farcs5$\times$4\arcsec\ &  1.7 \\
SBS 0335$-$052E& 20 Nov. 2009 & 1200                              &1\farcs5$\times$4\arcsec\ &  1.3 \\
  \enddata
\tablenotetext{a}{Extraction aperture for the one-dimensional spectrum.}
  \end{deluxetable}


  \begin{deluxetable}{lrrrr}
  \tabletypesize{\small}
  \tablecolumns{2}
  \tablewidth{0pc}
  \tablecaption{Fluxes of optical and NIR emission lines in II Zw 40
\label{tab3}}
  \tablehead{
 &&&&\multicolumn{1}{c}{CLOUDY} \\ \cline{5-5}
  \colhead{\sc Ion}
  &\colhead{100$\times$$F$($\lambda$)/$F$(H$\beta$)\tablenotemark{a}}
&\colhead{100$\times$$I$($\lambda$)/$I$(H$\beta$)\tablenotemark{b}}
&\colhead{100$\times$$I$($\lambda$)/$I$(H$\beta$)\tablenotemark{c}}
&\colhead{stars}
}
  \startdata
\multicolumn{5}{c}{a) Optical spectrum} \\
0.373 [O {\sc  ii}]                 &   23.87$\pm$0.66 &   59.06$\pm$2.03&   55.59$\pm$2.03& 53.28 \\
0.380 H10                           &    3.30$\pm$0.39 &    7.53$\pm$1.29&    7.49$\pm$1.29&  5.68 \\
0.384 H9                            &    3.91$\pm$0.34 &    8.70$\pm$1.27&    8.67$\pm$1.28&  7.70 \\
0.387 [Ne {\sc  iii}]               &   28.90$\pm$0.56 &   61.52$\pm$1.94&   62.24$\pm$1.94& 60.73 \\
0.389 He {\sc i}+H8                 &    8.52$\pm$0.34 &   18.14$\pm$1.39&   18.22$\pm$1.39& 17.84 \\
0.389 [Ne {\sc iii}]+H7             &   17.70$\pm$0.41 &   35.54$\pm$1.56&   35.89$\pm$1.57& 34.62 \\
0.410 H$\delta$                     &   17.89$\pm$0.37 &   32.61$\pm$1.36&   32.04$\pm$1.33& 26.34 \\
\enddata
\tablenotetext{a}{$F$(H$\beta$)=(1.097$\pm$0.004)$\times$10$^{-13}$ erg s$^{-1}$ cm$^{-2}$.}
\tablenotetext{b}{Corrected for extinction with $C$(H$\beta$)=1.165 and EW(abs)=0.250\AA.}
\tablenotetext{c}{Corrected for extinction with $C$(H$\beta$)=1.135 and EW(abs)=0.100\AA.}
  \end{deluxetable}


  \begin{deluxetable}{lrrrrr}
  \tabletypesize{\small}
  \tablecolumns{2}
  \tablewidth{0pc}
  \tablecaption{Fluxes of optical and NIR emission lines in Mrk 71
\label{tab4}}
  \tablehead{
 &&&&\multicolumn{2}{c}{CLOUDY} \\ \cline{5-6}
  \colhead{\sc Ion}
  &\colhead{100$\times$$F$($\lambda$)/$F$(H$\beta$)\tablenotemark{a}}
&\colhead{100$\times$$I$($\lambda$)/$I$(H$\beta$)\tablenotemark{b}}
&\colhead{100$\times$$I$($\lambda$)/$I$(H$\beta$)\tablenotemark{c}}
&\colhead{stars}&\colhead{stars+shock}
}
  \startdata
\multicolumn{5}{c}{a) Optical spectrum} \\
0.373 [O {\sc  ii}]                 &   37.61$\pm$0.56 &   44.09$\pm$1.11&   44.53$\pm$1.11& 35.52& 36.54 \\
0.375 H12                           &    2.58$\pm$0.11 &    3.35$\pm$0.30&    3.27$\pm$0.29&  3.54&  3.53 \\
0.377 H11                           &    3.02$\pm$0.11 &    3.84$\pm$0.32&    3.77$\pm$0.31&  4.38&  4.37 \\
0.380 H10                           &    3.98$\pm$0.11 &    4.95$\pm$0.35&    4.89$\pm$0.35&  5.68&  5.67 \\
0.382 He {\sc i}                    &    0.79$\pm$0.09 &    0.91$\pm$0.16&    0.92$\pm$0.16&  1.04&  1.03 \\
0.384 H9                            &    6.06$\pm$0.13 &    7.35$\pm$0.43&    7.32$\pm$0.42&  7.69&  7.67 \\
0.387 [Ne {\sc  iii}]               &   52.45$\pm$0.76 &   60.55$\pm$1.32&   61.27$\pm$1.32& 60.40& 60.47 \\
\enddata
\tablenotetext{a}{$F$(H$\beta$)=(4.112$\pm$0.004)$\times$10$^{-13}$ erg s$^{-1}$ cm$^{-2}$.}
\tablenotetext{b}{Corrected for extinction with $C$(H$\beta$)=0.225 and EW(abs)=0.6\AA.}
\tablenotetext{c}{Corrected for extinction with $C$(H$\beta$)=0.195 and EW(abs)=0.4\AA.}
  \end{deluxetable}


  \begin{deluxetable}{lrrrrr}
  \tabletypesize{\small}
  \tablecolumns{2}
  \tablewidth{0pc}
  \tablecaption{Fluxes of optical and NIR emission lines in Mrk 930
\label{tab5}}
  \tablehead{
 &&&&\multicolumn{2}{c}{CLOUDY} \\ \cline{5-6}
  \colhead{\sc Ion}
  &\colhead{100$\times$$F$($\lambda$)/$F$(H$\beta$)\tablenotemark{a}}
&\colhead{100$\times$$I$($\lambda$)/$I$(H$\beta$)\tablenotemark{b}}
&\colhead{100$\times$$I$($\lambda$)/$I$(H$\beta$)\tablenotemark{c}}
&\colhead{stars}&\colhead{stars+shock}
}
  \startdata
\multicolumn{4}{c}{a) Optical spectrum} \\
0.373 [O {\sc  ii}]                 &  207.01$\pm$3.37 &  250.96$\pm$4.94&  249.57$\pm$4.97&259.92&261.91 \\
0.384 H9                            &    3.66$\pm$0.31 &    9.16$\pm$0.92&   10.30$\pm$1.04&  7.70&  7.68 \\
0.387 [Ne {\sc  iii}]               &   29.83$\pm$0.70 &   35.40$\pm$1.29&   35.30$\pm$1.30& 38.61& 39.17 \\
0.389 He {\sc i}+H8                 &   12.66$\pm$0.46 &   20.00$\pm$1.03&   21.19$\pm$1.07& 21.48& 21.40 \\
0.389 [Ne {\sc iii}]+H7             &   20.54$\pm$0.58 &   28.98$\pm$1.19&   30.20$\pm$1.23& 27.94& 28.12 \\
0.407 [S {\sc ii}]                  &    1.92$\pm$0.46 &    2.20$\pm$0.56&    2.21$\pm$0.56&  1.45&  1.63 \\
0.410 H$\delta$                     &   19.58$\pm$0.55 &   26.82$\pm$1.12&   27.19$\pm$1.14& 26.29& 26.24 \\
\enddata
\tablenotetext{a}{$F$(H$\beta$)=(6.937$\pm$0.051)$\times$10$^{-14}$ erg s$^{-1}$ cm$^{-2}$.}
\tablenotetext{b}{Corrected for extinction with $C$(H$\beta$)=0.310 and EW(abs)=2.85\AA.}
\tablenotetext{c}{Corrected for extinction with $C$(H$\beta$)=0.270 and EW(abs)=3.55\AA.}
  \end{deluxetable}


  \begin{deluxetable}{lrrrr}
  \tabletypesize{\small}
  \tablecolumns{2}
  \tablewidth{0pc}
  \tablecaption{Fluxes of optical and NIR emission lines in Mrk 996
\label{tab6}}
  \tablehead{
 &&&&\multicolumn{1}{c}{CLOUDY} \\ \cline{5-5}
  \colhead{\sc Ion}
  &\colhead{100$\times$$F$($\lambda$)/$F$(H$\beta$)\tablenotemark{a}}
&\colhead{100$\times$$I$($\lambda$)/$I$(H$\beta$)\tablenotemark{b}}
&\colhead{100$\times$$I$($\lambda$)/$I$(H$\beta$)\tablenotemark{c}}
&\colhead{stars}
}
  \startdata
\multicolumn{5}{c}{a) Optical spectrum} \\
0.373 [O {\sc  ii}]                 &  149.06$\pm$2.33 &  198.11$\pm$3.78&  199.88$\pm$3.80&157.30 \\
0.377 H11                           &    0.83$\pm$0.29 &    6.01$\pm$2.77&    6.07$\pm$2.80&  4.64 \\
0.380 H10                           &    1.54$\pm$0.34 &    7.14$\pm$2.04&    7.21$\pm$2.06&  5.95 \\
0.384 H9                            &    2.59$\pm$0.28 &    8.37$\pm$1.21&    8.46$\pm$1.22&  7.98 \\
0.387 [Ne {\sc  iii}]               &   38.04$\pm$0.73 &   49.01$\pm$1.41&   49.55$\pm$1.41& 44.15 \\
0.389 He {\sc i}+H8                 &   11.82$\pm$0.38 &   20.60$\pm$0.98&   20.84$\pm$0.99& 19.70 \\
0.389 [Ne {\sc iii}]+H7             &   18.47$\pm$0.45 &   28.33$\pm$1.08&   28.69$\pm$1.08& 29.87 \\
\enddata
\tablenotetext{a}{$F$(H$\beta$)=(7.241$\pm$0.023)$\times$10$^{-14}$ erg s$^{-1}$ cm$^{-2}$.}
\tablenotetext{b}{Corrected for extinction with $C$(H$\beta$)=0.450 and EW(abs)=2\AA.}
\tablenotetext{c}{Corrected for extinction with $C$(H$\beta$)=0.420 and EW(abs)=2\AA.}
  \end{deluxetable}


  \begin{deluxetable}{lrrrrr}
  \tabletypesize{\small}
  \tablecolumns{2}
  \tablewidth{0pc}
  \tablecaption{Fluxes of optical and NIR emission lines in SBS 0335--052E
\label{tab7}}
  \tablehead{
 &&&&\multicolumn{2}{c}{CLOUDY} \\ \cline{5-6}
  \colhead{\sc Ion}
  &\colhead{100$\times$$F$($\lambda$)/$F$(H$\beta$)\tablenotemark{a}}
&\colhead{100$\times$$I$($\lambda$)/$I$(H$\beta$)\tablenotemark{b}}
&\colhead{100$\times$$I$($\lambda$)/$I$(H$\beta$)\tablenotemark{c}}
&\colhead{stars}&\colhead{stars+shock}
}
  \startdata
\multicolumn{5}{c}{a) Optical spectrum} \\
0.373 [O {\sc  ii}]                 &   28.34$\pm$0.47 &   31.11$\pm$0.91&   31.43$\pm$0.90& 27.36& 27.17 \\
0.375 H12                           &    2.25$\pm$0.18 &    2.94$\pm$0.39&    2.83$\pm$0.39&  3.43&  3.38 \\
0.377 H11                           &    3.85$\pm$0.16 &    5.10$\pm$0.60&    4.89$\pm$0.60&  4.26&  4.20 \\
0.380 H10                           &    4.99$\pm$0.17 &    6.15$\pm$0.52&    6.01$\pm$0.52&  5.52&  5.44 \\
0.384 H9                            &    7.61$\pm$0.18 &    8.98$\pm$0.56&    8.88$\pm$0.56&  7.49&  7.38 \\
0.387 [Ne {\sc  iii}]               &   22.27$\pm$0.36 &   24.23$\pm$0.78&   24.52$\pm$0.78& 24.66& 23.73 \\
0.389 He {\sc i}+H8                 &   16.03$\pm$0.28 &   17.94$\pm$0.69&   18.00$\pm$0.68& 19.11& 18.80 \\
\enddata
\tablenotetext{a}{$F$(H$\beta$)=(6.672$\pm$0.012)$\times$10$^{-14}$ erg s$^{-1}$ cm$^{-2}$.}
\tablenotetext{b}{Corrected for extinction with $C$(H$\beta$)=0.135 and EW(abs)=0.5\AA.}
\tablenotetext{c}{Corrected for extinction with $C$(H$\beta$)=0.105 and EW(abs)=0.35\AA.}
  \end{deluxetable}


\begin{deluxetable}{lrrrrrr}
  \tabletypesize{\small}
  \tablecolumns{7}
  \tablewidth{0pt}
  \tablecaption{Extinction-corrected fluxes of strong hydrogen lines 
\label{tab8}}
  \tablehead{
 &\multicolumn{5}{c}{Galaxy} & \\ \cline{2-6}
\colhead{Line}&\colhead{II Zw 40}&\colhead{Mrk 71}&\colhead{Mrk 930}&
\colhead{Mrk 996}&\colhead{SBS 0335$-$052E}
&\colhead{Theor.\tablenotemark{a}}
}
  \startdata
& \multicolumn{6}{c}{a) Optical spectrum} \\
0.410 H$\delta$      &  32.0 &  26.8 &  26.8 &  29.5 &  25.8 &  26.2 \\
0.434 H$\gamma$      &  48.9 &  48.7 &  46.3 &  54.9 &  47.8 &  47.3 \\
0.486 H$\beta$       & 100.0 & 100.0 & 100.0 & 100.0 & 100.0 & 100.0 \\
0.656 H$\alpha$      & 283.9 & 278.7 & 284.1 & 291.2 & 273.8 & 279.0 \\
& \multicolumn{6}{c}{b) NIR spectrum} \\
1.094 P$\gamma$      &   7.9 &   9.4 &   9.2 &   8.3 &   8.1 &   8.6 \\
1.282 P$\beta$       &  13.2 &  15.9 &  15.8 &  14.8 &  13.5 &  15.2 \\
2.165 Br$\gamma$     &   2.8 &   2.4 &   2.5 &   2.7 &   3.0 &   2.5 \\
  \enddata
\tablenotetext{a}{Recombination ratios for $T_e$=15000 K and 
$N_e$ = 100 cm$^{-3}$ from \citet{HS87}.}
  \end{deluxetable}


\begin{deluxetable}{lccccc}
  \tabletypesize{\small}
  \tablecolumns{6}
  \tablewidth{0pt}
  \tablecaption{Extinction coefficients $C$(H$\beta$) derived from individual hydrogen
line to H$\beta$ flux ratios\tablenotemark{a}
\label{tab9}}
  \tablehead{
 &\multicolumn{5}{c}{Galaxy} \\ \cline{2-6}
\colhead{Line}&\colhead{II Zw 40}&\colhead{Mrk 71}&\colhead{Mrk 930}&
\colhead{Mrk 996}&\colhead{SBS 0335$-$052E}
}
  \startdata
& \multicolumn{5}{c}{a) Optical spectrum} \\
0.410 H$\delta$      &0.686$\pm$0.215&0.112$\pm$0.150&0.348$\pm$0.287&0.188$\pm$0.224&0.051$\pm$0.159 \\
0.434 H$\gamma$      &1.001$\pm$0.247&0.091$\pm$0.216&0.416$\pm$0.317&0.005$\pm$0.244&0.026$\pm$0.221 \\
0.656 H$\alpha$      &1.181$\pm$0.115&0.217$\pm$0.112&0.348$\pm$0.129&0.403$\pm$0.114&0.093$\pm$0.112 \\
& \multicolumn{5}{c}{b) NIR spectrum} \\
1.094 P$\gamma$      &1.066$\pm$0.180&0.231$\pm$0.180&0.341$\pm$0.069&0.381$\pm$0.184&0.048$\pm$0.217 \\
1.282 P$\beta$       &1.036$\pm$0.164&0.200$\pm$0.163&0.331$\pm$0.052&0.387$\pm$0.166&0.016$\pm$0.197 \\
2.165 Br$\gamma$     &1.167$\pm$0.139&0.163$\pm$0.139&0.305$\pm$0.136&0.443$\pm$0.141&0.175$\pm$0.168 \\
  \enddata
\tablenotetext{a}{The adopted theoretical recombination ratios 
are for $T_e$=15000 K and 
$N_e$ = 100 cm$^{-3}$ from \citet{HS87} (see Table \ref{tab8}).}
  \end{deluxetable}


\begin{deluxetable}{lcccccccc}
  \tablecolumns{9}
  \tablewidth{0pt}
  \tabletypesize{\small}
  \tablecaption{H$_2$ emission-line flux ratios \label{tab10}}
  \tablehead{
 & \multicolumn{5}{c}{Galaxy} && \multicolumn{2}{c}{Model\tablenotemark{a}} \\ 
\cline{2-6} \cline{8-9}
\colhead{Line}&\colhead{II Zw 40}&\colhead{Mrk 71}&\colhead{Mrk 930}
&\colhead{Mrk 996}&\colhead{SBS 0335$-$052E}&&\colhead{Fluor}&\colhead{Coll}
}
  \startdata
1.162 H$_2$ 2-0 S(1)&\nodata    &0.1$\pm$0.1&\nodata    &\nodata    &\nodata    &&0.4&0.0 \\
1.207 H$_2$ 3-1 S(2)&0.2$\pm$0.1&0.2$\pm$0.1&\nodata    &\nodata    &\nodata    &&0.2&0.0 \\
1.233 H$_2$ 3-1 S(1)&0.4$\pm$0.2&0.4$\pm$0.2&\nodata    &\nodata    &\nodata    &&0.5&0.0 \\
1.238 H$_2$ 2-0 Q(1)&0.6$\pm$0.3&0.3$\pm$0.2&\nodata    &\nodata    &\nodata    &&0.4&0.0 \\
1.311 H$_2$ 4-2 S(1)&0.4$\pm$0.2&0.3$\pm$0.2&\nodata    &\nodata    &\nodata    &&0.4&0.0 \\
1.314 H$_2$ 3-1 Q(1)&0.4$\pm$0.2&0.3$\pm$0.2&\nodata    &\nodata    &\nodata    &&0.6&0.0 \\
1.601 H$_2$ 6-4 Q(1)&0.3$\pm$0.1&0.1$\pm$0.1&\nodata    &\nodata    &\nodata    &&0.4&0.0 \\
1.957 H$_2$ 1-0 S(3)&0.7$\pm$0.2&0.9$\pm$0.3&\nodata    &0.4$\pm$0.1&\nodata    &&\nodata&\nodata \\
2.034 H$_2$ 1-0 S(2)&0.4$\pm$0.1&0.2$\pm$0.1&\nodata    &0.3$\pm$0.1&\nodata    &&0.5&0.3 \\
2.073 H$_2$ 2-1 S(3)&0.3$\pm$0.1&0.2$\pm$0.1&\nodata    &\nodata    &\nodata    &&0.2&0.0 \\
2.122 H$_2$ 1-0 S(1)&1.0$\pm$0.2&1.0$\pm$0.2&1.0$\pm$0.2&1.0$\pm$0.2&1.0$\pm$0.4&&1.0&1.0 \\
2.223 H$_2$ 1-0 S(0)&0.4$\pm$0.1&0.5$\pm$0.2&0.6$\pm$0.2&\nodata    &\nodata    &&0.6&0.3 \\
2.248 H$_2$ 2-1 S(1)&0.6$\pm$0.2&0.5$\pm$0.2&0.5$\pm$0.1&0.3$\pm$0.1&0.6$\pm$0.4&&0.5&0.0 \\
2.355 H$_2$ 2-1 S(0)&0.2$\pm$0.1&0.2$\pm$0.2&\nodata    &0.2$\pm$0.1&\nodata    &&0.3&0.0 \\
 \enddata
\tablenotetext{a}{Model values are from \citet{BD87}. We adopt their
model 1 for fluorescent lines and model S1 for collisionally excited lines.}
  \end{deluxetable}


\begin{deluxetable}{lccccc}
  \tablecolumns{6}
  \tablewidth{0pt}
\tabletypesize{\small}
  \tablecaption{Input parameters for stellar photoionization 
CLOUDY models \label{tab11}}
  \tablehead{
\colhead{Parameter}&\colhead{II Zw 40}&\colhead{Mrk 71}&\colhead{Mrk 930}
&\colhead{Mrk 996\tablenotemark{a}}&\colhead{SBS 0335$-$052E}
}
  \startdata
$Q$(H)                 & 53.136 & 52.064 & 53.893 & 52.750 & 53.562 \\
$T_{eff}$, K           & 50000  & 50000  & 50000  & 50000  & 50000  \\
$N_e$, cm$^{-3}$       & 100    & 100    &  10    &        & 100    \\
log $f$                & --1.10 & --0.30 & --2.30 & --4.58 & --1.60 \\
log He/H               & --1.07 & --1.07 & --1.05 & --0.93 & --1.09 \\
log C/H                & --4.68 & --4.68 & --4.30 & --4.47 & --5.57 \\
log N/H                & --5.29 & --5.83 & --5.34 & --4.90 & --6.13 \\
log O/H                & --4.00 & --4.15 & --3.92 & --3.97 & --4.66 \\
log Ne/H               & --4.71 & --4.87 & --4.70 & --4.67 & --5.40 \\
log Si/H               & --5.35 & --5.35 & --5.35 & --5.57 & --6.05 \\
log S/H                & --5.63 & --5.78 & --5.70 & --5.50 & --6.53 \\
log Cl/H               & --7.49 & --7.59 & --7.49 & --7.47 &\nodata \\
log Ar/H               & --6.20 & --6.41 & --6.37 & --6.34 & --7.06 \\
log Fe/H               & --5.80 & --5.90 & --5.99 & --5.70 & --6.10 \\
  \enddata
\tablenotetext{a}{For Mrk 996, a two-zone model for the electron number 
density spatial distribution 
has been adopted. For the inner high-excitation zone with radii $r$ 
between 0.8 pc and 80 pc, the electron number density has  
a power-law dependence $N_e$ $\sim$ $r^{-2}$, 
decreasing from 4.5$\times$10$^6$ cm$^{-3}$ 
to 450 cm$^{-3}$. For the lower-excitation outer zone with $r$$>$ 80 pc,
the electron number density has been set to a constant value of 450 cm$^{-3}$.}
  \end{deluxetable}


  \begin{deluxetable}{lccccc}
  \tabletypesize{\small}
  \tablecolumns{7}
  \tablewidth{0pc}
  \tablecaption{Fluxes of NIR [Fe {\sc ii}] emission lines relative to 
Br$\gamma$ \label{tab12}}
  \tablehead{
 &&&&\multicolumn{2}{c}{CLOUDY} \\ \cline{5-6}
  \colhead{\sc Ion}
  &\colhead{$F$($\lambda$)/$F$(Br$\gamma$)}
&\colhead{$I$($\lambda$)/$I$(Br$\gamma$)\tablenotemark{a}}
&\colhead{$I$($\lambda$)/$I$(Br$\gamma$)\tablenotemark{b}}
&\colhead{stars}&\colhead{stars+shock}
}
  \startdata
 & \multicolumn{5}{c}{a) II Zw 40} \\
1.257 [Fe {\sc ii}]                 &  0.076 & 0.112 & 0.109 & 0.149 &\nodata \\
1.643 [Fe {\sc ii}]                 &  0.073 & 0.088 & 0.087 & 0.146 &\nodata \\
 & \multicolumn{5}{c}{b) Mrk 71} \\
1.257 [Fe {\sc ii}]                 &  0.034 & 0.036 & 0.037 & 0.108 & 0.159 \\
 & \multicolumn{5}{c}{c) Mrk 930} \\
1.257 [Fe {\sc ii}]                 &  0.745 & 0.892 & 0.876 & 0.485 & 0.652 \\
1.643 [Fe {\sc ii}]                 &  0.806 & 0.788 & 0.772 & 0.597 & 0.633 \\
 & \multicolumn{5}{c}{d) Mrk 996} \\
1.257 [Fe {\sc ii}]                 &  0.378 & 0.411 & 0.407 & 0.973 &\nodata \\
1.643 [Fe {\sc ii}]                 &  0.376 & 0.374 & 0.370 & 0.938 &\nodata \\
 & \multicolumn{5}{c}{e) SBS 0335--052E} \\
1.643 [Fe {\sc ii}]                 &  0.075 & 0.077 & 0.077 & 0.121 & 0.347 \\
  \enddata
\tablenotetext{a}{Corrected for extinction derived from the observed
Balmer decrement.}
\tablenotetext{b}{Corrected for extinction derived from the Balmer decrement,
itself corrected for collisional excitation of hydrogen lines.}
\end{deluxetable}



 \begin{deluxetable}{lrrr}
  \tabletypesize{\small}
  \tablecolumns{4}
  \tablewidth{0pt}
  \tablecaption{Fluxes of MIR emission lines \label{tab13}}
  \tablehead{
 &&\multicolumn{2}{c}{CLOUDY} \\ \cline{3-4}
  \colhead{\sc Ion}
&\colhead{100$\times$$I$($\lambda$)/$I$(H$\beta$)}
&\colhead{stars}&\colhead{stars+shock}
}
  \startdata
 & \multicolumn{3}{c}{a) II Zw 40\tablenotemark{a,b}} \\
10.51 [S {\sc  iv}]                 &   40.77$\pm$0.88& 68.19&\nodata \\
12.81 [Ne {\sc  ii}]                &    1.31$\pm$0.07&  0.43&\nodata \\
15.55 [Ne {\sc  iii}]               &   24.99$\pm$0.66& 33.56&\nodata \\
18.71 [S {\sc  iii}]                &   11.25$\pm$0.44& 10.40&\nodata \\
25.99 [Fe {\sc  ii}]                &    0.33$\pm$0.09&  0.13&\nodata \\
 & \multicolumn{3}{c}{b) Mrk 930\tablenotemark{c,d}} \\
12.81 [Ne {\sc  ii}]                &    5.11$\pm$0.37&  3.14&  3.35 \\
\enddata
\tablenotetext{a}{Spitzer data from \citet{W06} and \citet{O06}.}
\tablenotetext{b}{Observed H$\beta$ flux of 3.45$\times$10$^{-13}$ erg s$^{-1}$ cm$^{-2}$ from \citet{L07}.}
\tablenotetext{c}{Spitzer data from \citet{O06}.}
\tablenotetext{d}{Observed H$\beta$ flux of (2.54$\pm$0.10)$\times$10$^{-13}$ erg s$^{-1}$ cm$^{-2}$ from \citet{MK06}.}
\tablenotetext{e}{Spitzer data from \citet{T08}.}
\tablenotetext{f}{Extinction-corrected H$\beta$ flux of 6.6$\times$10$^{-13}$ erg s$^{-1}$ cm$^{-2}$ from \citet{T08}.}
\tablenotetext{g}{Model D from \citet{T08}.}
\tablenotetext{h}{Spitzer data from \citet{H04}.}
\tablenotetext{i}{Observed fluxes normalized to the extinction-corrected H$\beta$ flux of (1.813$\pm$0.027)$\times$10$^{-13}$ erg s$^{-1}$ cm$^{-2}$
\citep{I06b}.}
\end{deluxetable}

\clearpage

\clearpage

\clearpage

\clearpage

\clearpage

\clearpage

\clearpage

\clearpage

\clearpage

\clearpage

\end{document}